\documentclass[showpacs,amsmath,amssymb,floatfix]{revtex4}
\usepackage{graphicx}
\usepackage{dcolumn}

\begin{document}

\title{Reaction $^{48}$Ca$+^{208}$Pb: the capture-fission cross-sections and
the mass-energy distributions of fragments above and deep below the Coulomb barrier}

\author{E.~V.~Prokhorova, E.~A.~Cherepanov, M.~G.~Itkis, N.~A.~Kondratiev,
E.~M.~Kozulin, L.~Krupa, Yu.~Ts.~Oganessian and I.~V.~Pokrovsky}

\address{Flerov Laboratory of Nuclear Reactions,
Joint Institute for Nuclear Research,\\
141980, Dubna, Moscow region, Russia}

\author{V.~V.~Pashkevich}

\address{Bogoliubov Laboratory of Theoretical Physics,
Joint Institute for Nuclear Research,\\
141980, Dubna, Moscow region, Russia}

\author{ A.~Ya.~Rusanov}

\address{ Institute of Nuclear Physics of the National Nuclear Centre of Kazakhstan, \\
480082 Alma-Ata, Kazakhstan}

\begin{abstract}
The capture-fission cross-sections in an energy range of $206-242$~MeV 
of $^{48}$Ca-projectiles and mass-energy distributions (MED) of reaction 
products in an energy range of $211-242$~MeV have been measured in 
the $^{48}$Ca$+^{208}$Pb reaction using the double-arm time-of-flight 
spectrometer CORSET.  The MEDs of fragments
for heated fission were shown to consist of two components. One
component, which is due to classical fusion-fission, is associated with the
symmetric fission of the $^{256}$No compound nucleus. The other
component, which appears as ''shoulders'', is associated with the 
quasi-fission (QF) process and
can be named ''quasi-fission shoulders''. Those quasi-fission shoulders enclose
light fragments whose masses are $\approx 60-90$~a.m.u. The total
kinetic energy ($TKE$) of the fragments that belong to the shoulders is higher
than the value expected for a classical fusion-fission (FF) process. We
have come to the conclusion that in quasi-fission, spherical shells
with $Z=28$ and $N=50$ play a great role. It has also been demonstrated
that the properties of the MEDs of fragments formally agree with the standard
hypothesis of two independent fission modes; in this case the modes are
normal fusion-fission and quasi-fission processes. In this work, a version of 
the model based on the Dinuclear
System (DNS) concept has been developed, which made it possible to describe 
the experimental energy dependences of the capture-fission cross-sections
and the $xn$-channels of the reaction. 
The model takes into account shell effects and qualitatively predicts  the mass 
distributions (MDs) for quasi-fission process. A high-energetic
Super-Short mode of classical fission has been found at low excitation
energies in the mass range of heavy fragments $M = 130-135$ and
$ TKE \approx 233$~MeV; however the yield associated with this mode is small.
Its properties can be explained in the framework of the classical modal approach 
based on the valley structure of the potential energy surface in the deformation 
space of the fissioning nucleus. 

\end{abstract}

\date{\today}
\maketitle

\section{\label{sec:intro} INTRODUCTION}

In recent several years, considerable progress has been made in experiments on the synthesis
of superheavy elements. New isotopes of the nuclei with $Z = 102-116$ were discovered
(for example [1-3]). Of particular interest are fusion reactions of $^{48}$Ca ions with
various heavy targets because of the possibility of producing neutron-rich superheavy
nuclei, which, according to theoretical calculations [4], are the most stable nuclei.
In three such reactions, namely $^{48}$Ca($^{238}$U,$3n$)$^{283}112$, 
$^{48}$Ca($^{244}$Pu,$3n)^{289}114$, $^{48}$Ca($^{244}$Pu,$4n)^{288}114$, 
and  $^{48}$Ca($^{248}$Cm,$4n)^{292}116$
long-lived isotopes were produced of element $112$ and, for the
first time, of elements $114$ and $116$ [2, 3].

However, as known, the excited superheavy systems produced in reactions with massive
heavy ions undergo fission or competitive quasi-fission, which are their dominant decay
channels and whose characteristics carry information on collision dynamics, the life and
decay of the composite nucleus. The relation between the contributions of fission and
quasi-fission to the total ion-target interaction cross-section depends quite strongly
on the charges $Z_i$, $Z_t$ and masses $A_i$, $A_t$ of the ion
and target involved. For example, for
the reaction $^{48}$Ca$+^{208}$Pb, the fission of the compound nucleus
of $^{256}$No [5] is the main
decay process, whereas in the reaction $^{48}$Ca$+^{238}$U, quasi-fission,
a process in which no
classical compound nucleus is formed, dominates [6-8].

We studied both of those reactions for the energies of bombarding
$^{48}$Ca-ions near and below the Coulomb barrier of interacting partners.
In this work, we report on the
experimental results obtained for the reaction $^{48}$Ca$+^{208}$Pb:
on the capture-fission cross-sections and
structural peculiarities of the mass-energy distributions (MEDs) of the fission fragments
of the excited compound nucleus $^{256}$No. We also proposed a version of the
model based on the Dinuclear System concept, which describes competition
between fusion-fission and quasi-fission processes.

Recently, that reaction has already been studied in a few works: in [5, 9, 10],
the capture-fission cross-sections were measured, and, in [5, 11], the MEDs of
fragments. Our interest in it is connected with the fact that the possibility has
emerged of investigating the characteristics of that reaction deep below the Coulomb
barrier, where a compound system of quite low excitation energy (cold fusion
reactions [12]) can be produced, which is unattainable in other ion-target combinations.
Such experiments were made possible by modernisation of the FLNR U-$400$ accelerator ion
source, which enabled the production of $^{48}$Ca-ion beams of high quality and required
energy [13], and by development of the double-arm time-of-flight wide-aperture spectrometer
CORSET [14], with whose help the velocities and coordinates of nuclear reaction products
produced with record low cross-sections can be measured with a sufficiently high
resolution [15]. Some preliminary results of those experiments were presented in brief
reports [16, 17].

\section{ \label{sec:measuring_tech} Measuring technique and data processing}

The experiment was carried out on the extracted beam of the U-400 accelerator at the
Flerov Laboratory of Nuclear Reactions, Dubna. The capture-fission cross-sections were
measured in the range of the energies of bombarding $^{48}$Ca ions from $206$ to $242$~MeV;
the MEDs of reaction products were measured in the range $211-242$~MeV. The target
was a $220~\mu$g/cm$^2$ $^{208}$Pb layer deposited on a $50~\mu$g/cm$^2$
carbon backing foil.

To register reaction products, the well known kinematic coincidence method [5-8, 10,18]
was applied with the use of the double-arm time-of-flight spectrometer CORSET [14].
Each arm of the CORSET spectrometer consisted of a start detector composed of microchannel
plates with an electrostatic mirror and two stop position-sensitive ($x,y$-sensitivity)
detectors also composed of  microchannel plates, $6 \times 4$~cm in size.
The start detectors
were positioned at a distance of $3.5$~cm from the target. The minimum start-stop flight
distance was $11.5$~cm. Thus, the spectrometer consisted of two start detectors and $4$ stop
detectors and covered a solid angle of $360$~msr. The position resolution of the stop
detectors was $0.7$~mm. The mass resolution of the spectrometer was checked against the
MEDs of fragments for the spontaneous fission of $^{252}$Cf; our estimate of it
is $3-5$~u. [19].
 Each arm of the spectrometer could be independently moved along the radius of the reaction
 chamber without vacuum failure, so that the arms could be installed at selected correlation angles.
 The technique used to measure the capture-fission cross-sections
for different correlation angles was the same as described in work [10]. To absolutely
normalise cross-sections and to monitor the ion beam, two semiconductor
monitors capable of registering elastically scattered ions of $^{48}$Ca were
installed at angles of $\pm 11^\circ$.

\begin{figure}[p]
\includegraphics[width=9cm]{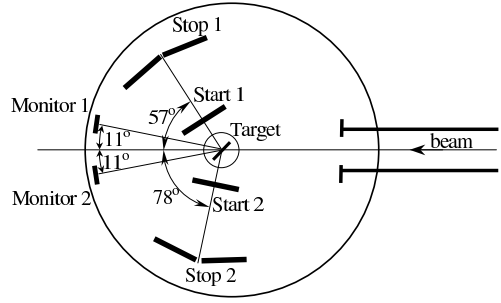}
\caption{\label{fig:fig1} Schematic diagram of
the experimental arrangement}
\end{figure}

\begin{figure}[p]
\includegraphics[width=7cm]{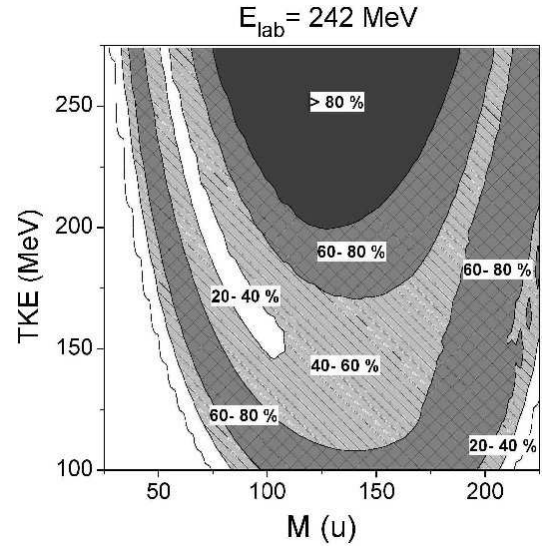}
\caption{\label{fig:fig2} Geometrical efficiency of the spectrometer (in \%) on $(M - TKE)$
coordinates.}
\end{figure}

To measure MEDs of reaction products, the spectrometer arms were positioned at
correlation angles of $57^\circ$ and $78^\circ$ correspondingly. The registering angles in
the laboratory system were $\theta _{\rm lab 1}= 39^\circ-75 ^\circ$
and $\theta _{\rm lab 2}=  61^\circ- 95^\circ$, which corresponds in center-of-mass system
for registerated products to $\theta _{\rm CM  1}= 47-113^\circ$,
$\theta _{\rm CM 2}= 73^\circ-140^\circ $.
In this case, the spectrometer was capable
of registering all reaction products with the mass of from a bombarding ion to target
nuclei. The experiment geometry is schematically depicted in Fig.~\ref{fig:fig1}. Since the
spectrometer acceptance did not allow the whole range of energy-mass ratios to
be registered with the same probability, corrections for its geometrical efficiency
were introduced. As an example, a two-dimensional map (in percent) of that efficiency
in mass-energy coordinates is presented in Fig.~\ref{fig:fig2} for the bombarding energy
$E_{\rm  lab}=242$~MeV. That map was  calculated by the Monte-Carlo method
proceeding from the two-body nature of the reaction exit channel.

The data were processed by the standard technique; for details, see
works [7, 8, 18].  Energy losses of reaction products in the target layer,
backing foil and in electron emitters of the start detectors were taken into account.
The spectrometer was calibrated in two stages. At the preliminary stage, the
spectrometer arms were positioned opposite each other and $^{252}$Cf fission fragments
were used to calibrate the spectrometer. At the final stage, the arms were in the
operating position and the calibration was carried out of the energy 
of registered elastically scattered
ions of $^{48}$Ca and $^{208}$Pb. Particular attention was given to the angular folding correlation of
products both in and out of the reaction plane. Only those events which correspond to 
the two-body process of complete momentum transfer were selected and then analysed.
Let us give a typical example of such selection.

\begin{figure}[p]
\includegraphics[height=7 cm]{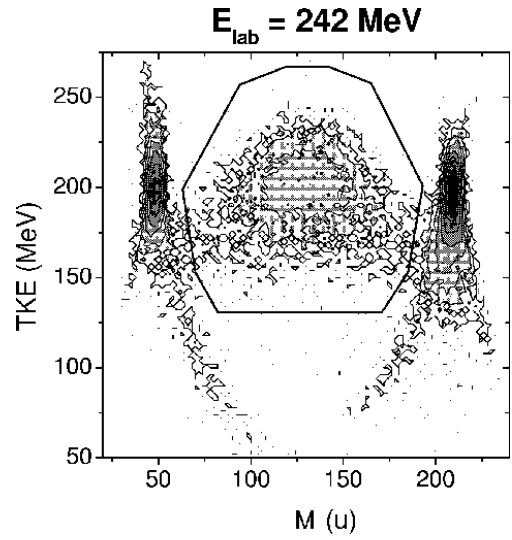}
\caption{\label{fig:fig3} Two dimensional $(M-TKE)$ matrix for the beam energy $E_{\rm lab}=242$~MeV 
without two-body events selection.}
\end{figure}

\begin{figure}[p]
\includegraphics[angle=-90, width=12cm]{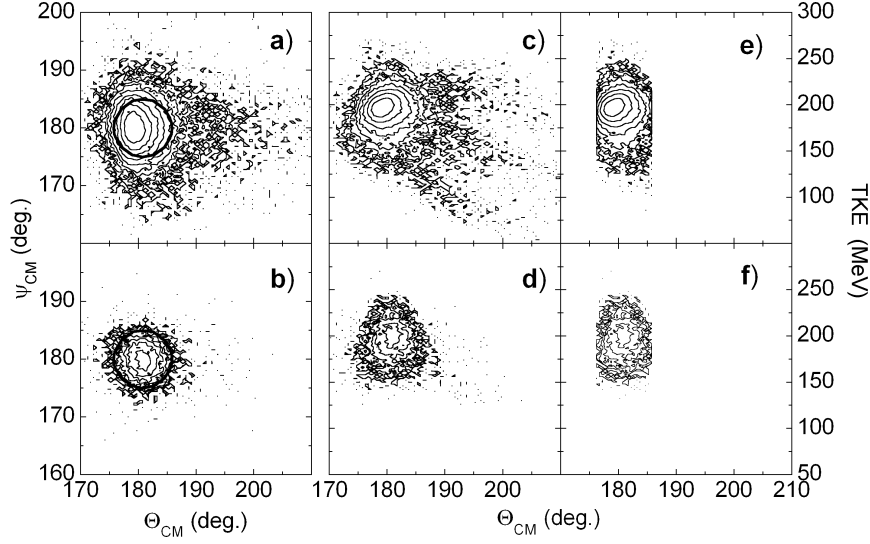}
\caption{\label{fig:fig4} a) and b) Two-dimensional matrixes for the sum 
of the center-of-mass angles of divergence of reaction products; 
$\Theta_{\rm CM}$ --  in and  $\Psi_{\rm CM}$ -- out-of-plane: a) for all the events registered, b) for
fission fragments only. Only the events enclosed by curves were
analyzed. Figs. c) and d) two-dimensional $\Theta_{\rm CM}-TKE$ matrixes for
reaction products; c) for all the events, d) for fragments only. Figs.
e) and f) the same as in Figs. c) and d) but only for the events
enclosed by curves in Figs. a) and b).}
\end{figure}

Figure~\ref{fig:fig3} shows the two-dimensional matrix ($M - TKE$) (mass of reaction products and
their total kinetic energy) for all the registered events, that is, summed over all
the measured angles, for the energy of $^{48}$Ca bombarding ions $E_{\rm  lab}=242$~MeV on the
assumption that all the events are two-body events. Fig.~\ref{fig:fig4} presents the two-dimensional
matrices for those events as a function of the sum of the angles in the
center-of-mass system in the reaction plane $\Theta _{\rm CM}$ and
out-of-plane $\Psi _{\rm CM}$; Fig.~\ref{fig:fig4}(a) presents the matrices for all the
events; Fig.~\ref{fig:fig4}(b) only for the fission fragments extracted from the matrix ($M - TKE$)
as shown in Fig.~\ref{fig:fig3}. It is clearly visible that for fragments (Fig.~\ref{fig:fig4}(b)),
matrix ($\Theta _{\rm CM} - \Psi _{\rm CM}$)
is quite symmetric around $180^\circ$ in both planes, only few events are located to the
right from it, which points to fission occurring after incomplete momentum transfer.
We only took account of the events lying within the contours indicated in Figs.~\ref{fig:fig4}(a)
and \ref{fig:fig4}(b).
In Fig.~\ref{fig:fig4}(a), there is a different situation for all the registered events. It is
clearly visible that the distribution is extended toward the large angles in the
reaction plane, which is evidence of a quite major contribution due to a
non-two-body process that does not result to the capture and fission in the
peripheral interaction of colliding partners. 
In Figs.~\ref{fig:fig4}(c), (d), (e) and (f), the same data
are represented in a more illustrative way and in the ($\Theta _{\rm CM} - TKE$) coordinates;
Fig.~\ref{fig:fig4}(c)
presents all the registered event; Fig.~\ref{fig:fig4}(d) only shows the fragments.
In Figs.~\ref{fig:fig4}(e) and \ref{fig:fig4}(f),
the selected events from within the contours in Figs.~\ref{fig:fig4}(a),(b) are shown in the same
coordinates. Later on, the events confined in the contours in Figs.~\ref{fig:fig4}(a),(b) 
were only analysed. The same procedure was applied for the other energies of $^{48}$Ca-projectiles.

Thus we managed to separate quite reliably the two-body process of system decay from 
the background of all other reaction channels and then we could extract  physical 
information only from the events, corresponding to the above mentioned criteria.

\section{ Results}
\label{sec:res}

Table~\ref{tab:table1} gives the parameters of the studied reactions and some measured integral
characteristics of fragments.

\begin{figure}[p]
\includegraphics[width=5cm]{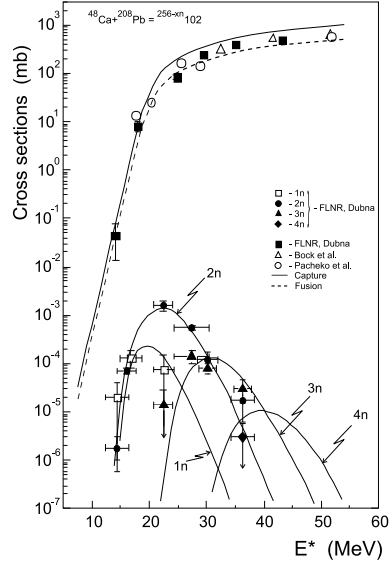}
\caption{\label{fig:fig5} The dependence of the capture-fission cross-sections 
$\sigma _{ c}$ on
excitation energy $E^*$. $\bullet$ - our data,
$\bigtriangleup$ -- data from ref. [5], $\circ$ -- data
from ref.[10]. Shown at the bottom are the data from work [20]on
the cross-sections of the $xn$-channels for the same reaction. The
capture-fission cross-sections $\sigma _{c}$ and
the cross-sections of the $xn$-channels calculated in the framework
of our version of the DNS
model are indicated with solid curves. The calculated complete
fusion cross-section $\sigma _{\rm CN}$ is indicated with a dashed curve.
}
\end{figure}

At the top of Fig.~\ref{fig:fig5}, the capture-fission cross-section is shown as a function of
the excitation energy of the compound nucleus $E^*$. Our results are in good agreement
with the data from works [5,10], which are also shown in Fig.~\ref{fig:fig5}. At
the bottom of that
figure, the data are presented on the cross-sections of the $xn$-channels of the reaction
under study from work [20]. With solid curves, the description is shown of those data on
the basis of the Dinuclear System Model, which is to be discussed in detail later on.
The dotted curve represents the cross-section for the complete
fusion of $^{48}$Ca and $^{208}$Pb
ions also calculated on the basis of that model.

\begin{figure}[p]
\includegraphics[angle=-90,width=12cm]{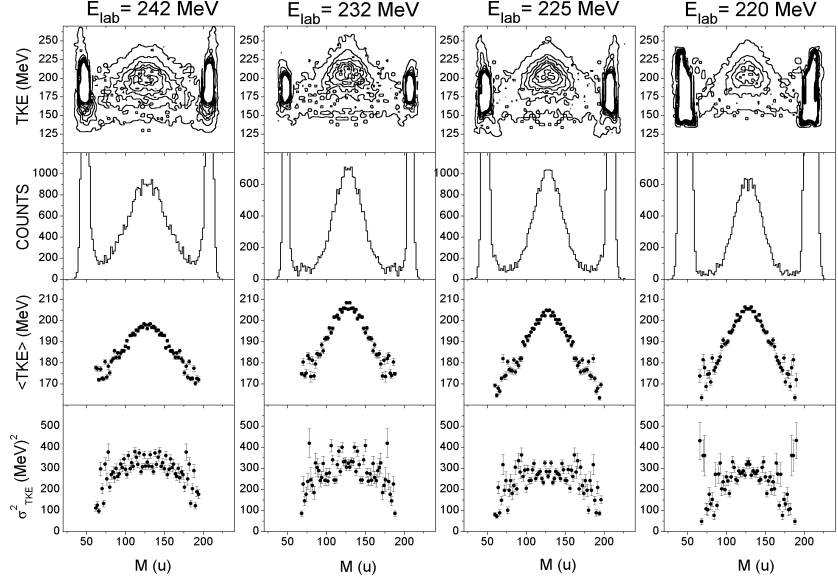}
\caption{\label{fig:fig6} Shown from top to bottom for the indicated energies
of $^{48}$Ca-projectiles are: the two-dimensional ($M-TKE$) reaction product
matrixes, the mass distributions and, for fission fragments only,
the dependences of $\langle TKE \rangle$ and its
variance $\sigma ^2 _{\rm TKE}$ on fragment mass $M$.}
\end{figure}

Figures~\ref{fig:fig6}, \ref{fig:fig7} and \ref{fig:fig8} demonstrate the MEDs of reaction
products for all the studied energies
of $^{48}$Ca ions. Shown from top to down in those figures are: the two-dimensional matrices
($M - TKE$), mass distributions and next, only for fragments, the $TKE$ and its variance
$\sigma^2_{\rm TKE}$ as functions of fragment mass.

Figure~\ref{fig:fig6} gives the measured results for the MEDs
of fragments for the four largest values
of the energy of $^{48}$Ca-ions. The reaction products near the
masses of bombarding ion and
target ($\cong M_{i, t} \pm 20$) are easy to identify as elastic,
quasi-elastic and deep inelastic
events. At the same time, it is clearly visible that the
fragment mass distributions for
symmetric masses with $A/2 \pm  30$ are of a shape close to a
gaussoid and that the dependences
of $\langle TKE \rangle(M)$ and $\sigma^2_{\rm TKE}(M)$ are of  parabolic shapes.
It is this behaviour of the indicated
characteristics of fragments
that is typical of the fission of heated nuclei,
as predicted by the liquid-drop model (LDM) [21] or the diffusion model [22], since
the shell properties of a fissioning nucleus become not essential, that is, their
distinctive features are inherent in standard classical fission.
At the same time, for
light fragments with masses $65 \leq M \leq 100$ and the masses
complementary to those values,
the mass distributions show "shoulders"; those "shoulders" are
due to fact that the yield of reaction products
increases in comparison with that predicted for
classical fission. Each of those events
can hardly be thought to be a deep inelastic process as well because they belong in the
matrix of fragments, that is, they are fragment-like, although to the right and the
left of that region, smooth transition is seen to events of deep inelastic transfer
and to standard fission, and, on the whole, it is impossible to conclusively indicate
the boundary of ''shoulders''. It should be noted
that in Fig.~\ref{fig:fig6}, the ''shoulders'' are sufficiently
easy to separate from the main symmetric peak of fragments for the ion energy
$E_{\rm  lab}\leq 232$~MeV but no longer easy to separate for
the energy $E_{\rm  lab}=242$~MeV and  
one can judge about their presence from the sharp widening  of the MD bottom. 
In Fig.~\ref{fig:fig6}, in the region of fragment 
masses in which the ''shoulders'' dominate 
an increased $TKE$ is seen for all the four ion energies as compared with "normal" fission.
However, the data presented in Fig.~\ref{fig:fig6} are not statistically accurate
enough to state that the fragment
characteristics for the masses in the region of ''shoulders'' are different from classical
fission. To clarify this issue, which, in our opinion, is a principal one, we once
again measured the MEDs of fragments
for the energy of $^{48}$Ca ions $E_{\rm  lab}=230$~MeV
with statistics of events of about one order higher than those on the average in
Fig.~\ref{fig:fig6}. The results of that experiment are shown in Fig.~\ref{fig:fig7}.
It is clearly visible
that, indeed, for light fragments of the masses $M_L \leq 95$ and
the heavy masses complementary
to them, the $TKE$ obviously deviates upwards from the parabolic curve by which
the dependence $\langle TKE \rangle (M)$ is well described in the region of symmetric masses.
It is also seen that a small wide maximum, invisible in the same
distributions in Fig.~\ref{fig:fig6}, becomes faintly visible in the 
$\sigma^2_{\rm TKE}(M)$ dependence in the range of symmetric mass $M = 115-140$.
This means that though it is close to liquid-drop fission in properties, the
fission of $^{256}$No at $E_{\rm lab}=230$~MeV
($E^*=33$~MeV) has insignificant structural peculiarities characteristic of enough
low-energy  fission that do not manifest themselves directly in
the mass yields and the $TKE$ of symmetric fission fragments.

Recall that no ''shoulders'' are observed in the MEDs of the
compound nucleus of $^{255}$No produced in the reaction $^{16}$O$+^{239}$Pu,
which is also true for
the fissioning nuclei that are produced in reactions with ''light'' heavy nuclei, and
which are close to that nucleus.
In addition, the dependences $\langle TKE \rangle (M)$ and $\sigma^2_{\rm TKE}(M)$
show no peculiarities [23].
Then we will use the term ''quasi-fission shoulders'', without going into
detail yet, in reference to the light fragments with $M\leq 95$ and the
heavy masses complementary to them.

Figure~\ref{fig:fig8} gives the MEDs of fragments for the two lowest energies of $^{48}$Ca ions
$E_{\rm  lab}=217$ and $211$~MeV, which corresponds to the initial excitation energy of
the compound nucleus of $^{256}$No $E^*=22,5$ and $17,6$~MeV correspondingly. At those
$E^*$, according to systematics [24], practically no pre-fission neutrons are
emitted (see Table~\ref{tab:table1}) and, therefore, the actual temperature $ E^*$ with which the
nucleus breaks up is not different from the initial temperature. Those are rather
low excitation energies, and in Fig.~\ref{fig:fig8}, it is clearly visible that the shape of
the MD is far from a gaussoid: there are structured flat tops in the region of
the heavy masses $M_H \cong 128-135$, and at the lowest energy
($E_{\rm  lab} = 211$~MeV) there
is, in addition, a clearly defined increased yield of fragments for $M_H = 145-155$.
For the fragment masses near $A/2$ at $E_{\rm  lab}=217$~MeV,
in contrast to higher energies,
the distribution $\langle TKE \rangle (M)$ is obviously extended toward the larger $TKE$. That makes the
impression that the wide parabolic dependence $\langle TKE \rangle (M)$, characteristic of the fission
of more heated nuclei (Fig.~\ref{fig:fig6} and \ref{fig:fig7}), has a narrow more 
high-energetic component. For the
energy $E_{\rm  lab}=211$~MeV, only this narrow component basically remains, and here the
dependence $\sigma^2_{\rm TKE}(M)$ shows a statistically significant spike
in the variance. That
the MEDs of fragments presented in Fig.~\ref{fig:fig8} behave in such a manner points to the fact
that shell effects manifest themselves in the quite low-energy fission of the
compound nucleus of $^{256}$No.

\section{ Model}
\label{sec:mod}
\subsection{ Fundamentals of the Dinuclear System Concept}
\label{sec:dnsc}

The motivation for the DNS Concept (DNSC) and a comparison of the DNS Concept
with existing models for the fusion of massive nuclei have been already given
in works [25-28]. Therefore here we will only discuss its main features applied
to analysis of the fusion reactions used in the production of transfermium
elements. According to the DNS concept, complete fusion occurs as follows.
At the capture stage, after the kinetic energy is completely transferred into
the inner degrees of freedom, a dinuclear system is formed with the capture
cross-section $\sigma _c$. The capture cross-section $\sigma _c$
is part of the total inelastic
cross-section $\sigma _R$:
\begin{equation}
\label{eq:form1}
\frac{{\sigma _c }}{{\sigma _R }} = \left[ {\sum\limits_{l =
0}^{l_{cr} } {(2l + 1)T(l,E_{CM} )} } \right]/\left[
{\sum\limits_{l = 0}^\infty  {(2l + 1)T(l,E_{CM} )} } \right],
\end{equation}

where $l_{cr}$ is the critical angular momentum at which the heavy ion is captured
and the excited dinuclear system is formed. The value for $l_{cr}$ was taken from
empirical systematics for the $\sigma _c$/$\sigma _R$ ratio [29]. The capture
cross-section is
described by the formula:

\begin{equation}
\label{eq:form2}
\sigma _{c} (E^* ) = \pi \mathchar'26\mkern-10mu\lambda _0^2
\sum\limits_{l = 0}^{l_{cr} } {(2l + 1)T(l,E_{CM} )} ,
\end{equation}

where the penetration of the potential barrier is calculated on the basis of
the standard inverted parabola method and written as
\begin{equation}
\label{eq:form3}
T(l,E_{CM} ) = \left\{ {1 + \exp \left[ {(2\pi /\hbar w)(B - E)}
\right]} \right\}^{ - 1} ,
\mbox{ where  }
\hbar w = {\raise0.7ex\hbox{$\hbar $} \!\mathord{\left/
 {\vphantom {\hbar  {\sqrt \mu  }}}\right.\kern-\nulldelimiterspace}
\!\lower0.7ex\hbox{${\sqrt \mu  }$}}\left[
{{\raise0.7ex\hbox{${d^2 V(R,l)}$} \!\mathord{\left/
 {\vphantom {{d^2 V(R,l)} {dr^2 }}}\right.\kern-\nulldelimiterspace}
\!\lower0.7ex\hbox{${dr^2 }$}}} \right]^{1/2} ,
\end{equation}

Here $E_{CM}$ is the center-of-mass energy of the bombarding ion; $R$ -- the distance between the
centres of the nuclei; $\mu$ -- the reduced mass of the system; $B$ -- the Coulomb
barrier. In calculating the capture cross-section, the Woods-Saxon potential
was used as the interaction potential:
\begin{equation}
\label{eq:form4}
V_N (R) = \frac{{V_0 }}{{1 + \exp \left\{ {\left[ {R - r_0 \left(
{A_i^{1/3}  + A_t^{1/3} } \right)} \right]/d} \right\}}},
\end{equation}

whose parameters: $d$ -- the diffuseness of the potential, $V_0$ -- the depth of the
potential, and $r_0$ -- the radius parameter, were taken from work [29], in which
empirical systematics for the parameters of the potential had been found on
the basis of a large body of experimental data on the capture cross-section.

In our approach, complete fusion is the final stage of the evolution of a
DNS, at which all the nucleons of one nucleus will have gradually been
transferred to the other nucleus. In the best known models for nuclear fusion,
the production cross-section for compound nuclei, $\sigma_{\rm CN}$, is not different from the
capture cross-section, $\sigma _{c}$; in other words, after the capture stage, the compound
nucleus is formed with $100\%$ probability. In our approach, the complete fusion 
cross-section, $\sigma _{\rm CN}$, is part of the capture cross-section, $\sigma _{c}$.
The fusion process will compete with the quasi-fusion process, and therefore
the complete fusion cross-section can be written as
\begin{equation}
\sigma _{\rm CN} (E^* ) = \sigma _c  \cdot P_{\rm CN}  \approx \pi
\mathchar'26\mkern-10mu\lambda _0^2 \sum\limits_{l = 0}^{l_{cr} }
{(2l + 1)T(l,E_{CM} ) \cdot P_{\rm CN} } ,
\end{equation}

where $P_{\rm CN}$ is the probability that complete fusion will occur.

In the evolution of the DNS, each nucleus of the DNS retains its individuality;
this is a consequence of the influence of the shell structure of the partner nuclei
since the kinetic energy of the bombarding ion, and, thus, the resultant excitation
energy, as a rule, is low in those reactions. The Macroscopic Dynamic Model
(MDM) [30] description of the coalescence of two nuclear drops does not take
account of the shell structure of the nuclei, and complete fusion does not compete
with quasi-fission. Those two processes are considered to be distinctly
separated in energy space.

An essential characteristic of a DNS, determining its evolution, is the
system's potential energy $U(Z,J)$:
$
U(Z,J) = B_1 + B_2 + V(R^*, J) - [B_{CN} + V_{\rm rot}(J)],
$

where $Z$ -- charge of one of the DNS partners, $J$ -- spin of the system, which 
practically coincides with the angular moment of collision $L$ (further we shall 
use everywhere $L$); $B_1$, $B_2$ and $B_{CN}$ -- binding energies of DNS 
nucleus and a compound nucleus. $V(R^*,L)$ -- the nucleus-nucleus 
potential taken at point $R^*$ corresponding position of a minimum in potential 
energy, which includes Coulomb, nuclear and centrifugal potentials:

\begin{equation}
\label{eq:form6}
V(R,L) = V_C(R)+ V_N(R)+V_{\rm rot}(R,L).
\end{equation}

We took PROXIMITY (for details, see [31]) as the nuclear potential $V_N(R)$.

\begin{eqnarray}
V_N (R) = 4\pi \gamma  \cdot \overline R  \cdot b \cdot \Phi (\xi) & &\\
\mbox{ where } \gamma  = 0.9517 \cdot \left[ {1 - 1.7826\left( {\frac{{N -
Z}}{A}} \right)^2 } \right] \cdot MeV \cdot fm^2,\quad 
\overline R  = \frac{{C_P  \cdot C_T }}{{C_P  + C_T }}, \nonumber \\
\mbox{ and } C_i  = R_i (1 - b^2 /R_i ^2  + ....),\quad
\xi  = s/b, \quad
s = r - (C_P  + C_T ),\nonumber \\
R_i  = 1.28A_i^{1/3}  - 0.76 + 0.8A_i^{ - 1/3} ,\quad
b = 1.0, \quad
\mbox{$i=P$ or $T$ }\nonumber \\
\Phi (\xi ) = \left\{{\begin{array}{*{20}c}
   { - 1.7817 + 0.9270 \cdot \xi  + 0.14300 \cdot \xi ^2  - 0.09000 \cdot \xi ^3 ,}  \\
   { - 1.7817 + 0.9270 \cdot \xi  + 0.01696 \cdot \xi ^2  - 0.05148 \cdot \xi ,^3 }  \nonumber \\
   { - 4.41 \cdot \exp \left( { - \frac{\xi }{{0.7176}}} \right),}  \\
\end{array}} \right.\begin{array}{*{20}c}
  {\xi  < 0}  \\
  {0 < \xi  < 1.9475}  \\
  {\xi  > 1.9475}  \\
\end{array}
\end{eqnarray}

The Coulomb energy was calculated from the following formula:
\[
V_C  = Z_P Z_T e^2  \cdot \left\{ {\begin{array}{*{20}c}
   {1/r,} & {r \ge R_C }  \\
   {\frac{1}{{2R_C }}(3 - r^2 /R_C^2 ),} & {r < R_C }  \\
\end{array}} \right.,
\]
 where
$R_C  = r_C (A_P^{1/3}  + A_T^{1/3} )$

The rotation energy $V_{\rm rot}(R,L)$ was calculated with the solid moment of inertia.
Our estimates show that the rotation component has a small effect on the
calculation results of  competition between fusion and quasi-fission processes
for the ion energy close to the Coulomb barrier.

Shown in Fig.~\ref{fig:fig9} is a qualitative picture that presents three-dimensional PROXIMITY
potential as a function of the distance between the centres of the nuclei and the
charge number of one of the DNS fragments. On capturing a heavy ion by a target
nucleus, the formed DNS will fall into a potential pocket. Then the double
system will be moving along the valley of the potential pocket both in the direction
to symmetry and in the direction to decreasing the charge number of one of the
fragments --- in the direction to fusion. On the way, the DNS might undergo a
break-up into two fragments --- QF, i.e. fission without a compound nucleus being
produced. In the figure, the corresponding processes are indicated by arrows.
In Fig.~\ref{fig:fig10}, shown are two profiles of the potential energy: the profiles along
the minimum and the maximum of the potential surface calculated for the reaction
$^{48}$Ca$+^{208}$Pb. The curve $U(Z,L=0)$ (for the magnitude of $R$ corresponding to the pocket)
has a few local minima, which reflect the shell structure in interacting nuclei.
Heavy and superheavy elements (SHE) are typically produced at the ion energies
that result in  the minimally possible excitation energy of the compound nucleus.
This ensures higher survival probability for the compound nucleus while de-excited.
In Fig.~\ref{fig:fig10}, the DNS energy that corresponds to the minimally possible excitation
energy of the compound nucleus is shown by cross-hatching. As follows from Fig.~\ref{fig:fig10},
the system will undergo most heating while descending from the Businaro-Gallone
point (B.G.) to the point of compound nucleus formation. Only at this stage of
DNS evolution will most of the potential energy of the dinuclear system be
transformed into the heat excitation of the compound nucleus. This peculiarity
of DNS evolution, characteristic of SHE fusion reactions, required the use of
experimental masses to calculate the potential energy $U(Z,L)$. In addition, on
the way to the compound nucleus, the DNS should overcome the inner potential
barrier $B^{*}_{\rm fus}$, which is the difference between the magnitudes
of the potential energy $U(Z,L)$ at the B.G. and at the
reaction entrance point. The inner potential barrier $B^*_{\rm fus}$ results from nucleon
transfer in a massive DNS being of an endothermic nature, which makes the system
move in the direction to the compound nucleus. The movement of the DNS in the reverse
direction, to more symmetry, might result in its leaving the potential pocket (with a
break-up into two fragments --- the movement in the direction to increasing $R$) after
overcoming the QF barrier, which we took to be the difference between the magnitudes
of driving potential for the entrance combination and the point of scission into two
fragments. The required energy for overcoming these barriers is taken from the
excitation energy $E^*$ of the dinuclear system, which an essential feature of our
approach. A compound nucleus is unlikely to form if the DNS excitation energy is
smaller than the magnitude of $B^*_{\rm fus}$. The more symmetric the entrance combination
of nuclei, the higher the inner fusion barrier $B^*_{\rm fus}$, which the dinuclear system
has to overcome on the way to the compound nucleus, and the lower the quasi-fission
barrier $B_{\rm QF}$, hence QF offers stronger competition.

\subsection{ Competition between complete fusion and quasi-fission in
SHE synthesis reactions}
\label{sec:comp}

Nucleon transfer between the nuclei in a DNS has a statistic nature and there
is a chance that the system may reach and overcome the B. G. point, and
thus a compound nucleus will form. The alternative to that process will be the
break-up of the system into two fragments (quasi-fission process). To calculate the
probability of proton transfer from one nucleus to the other in a dinuclear system,
we used an expression from work [32] and assumed that the macroscopic nucleon transfer
probability $P_z$ can be expressed in terms of the microscopic probability
$\lambda _z$ and level density $\rho _z$ as $P_z = \lambda _z \cdot \rho _z$.
The level density can be written in terms of the DNS potential
energy as $\rho _z = \rho (E^*-U(Z,L))$, where $E^*$ is the excitation
energy of the double system.
Finally, the proton capture P$^+$ and delivery P$^-$ probabilities can be written as follows:
\begin{equation}
\label{eq:form8}
P^ +   = \left\{ {1 + \exp \left[ {\frac{{U(Z + 1,L) - U(Z -
1,L)}}{{2T}}} \right]} \right\}^{ - 1},\quad P^ -   = \left\{ {1 +
\exp \left[ {\frac{{U(Z - 1,L) - U(Z + 1,L)}}{{2T}}} \right]}
\right\}^{ - 1} ,
\end{equation}

where $T=\sqrt{(E^*/a)}$ is the nucleus temperature; $a=0.093A$ -- the level density parameter.
Knowing those relative ($P^+ + P^- =1$) probabilities and using a random value uniformly
distributed over the interval between 0 and 1, we randomly choose the direction for
the DNS to move in: either the direction to the symmetric system or the direction to
the compound nucleus, repeating this procedure as many times as needed to obtain the
necessary statistics.

Fig.~\ref{fig:fig11}  shows the calculated results --- the mass distribution of QF
products for the reaction $^{48}$Ca+$^{208}$Pb for $E^*=33$~MeV. It is
seen from that figure that the spectrum for
the mass distribution of reaction fragments correlates with the structure of the
driving potential. The maxima of the mass distribution are matched by the position
of the local minima of the potential. This is a reflection of the nature of our
approach to calculation in the case that there exist quasi-stationary states
(local minima), which gives grounds to apply the statistical model.

\subsection{   Survival probability for the statistical decay of an excited
compound nucleus}
\label{sec:survival}

The formula for the cross-section after the process of de-excitation of fissioning
excited compound nuclei can be written as follows:
\begin{equation}
\label{eq:form9}
\sigma _{ER} (E^* ) \approx \pi \mathchar'26\mkern-10mu\lambda ^2
\sum\limits_{l = 0}^{l_{cr} } {\sigma _{\rm CN} } (E^* ,l) \cdot
W_{\rm sur} (E^* ,l),
\end{equation}

where $\sigma _{\rm CN}$ is the production cross-section for the compound nucleus. There are a
few approaches to calculate the survival factor $W_{\rm sur}$ for the compound nucleus
in competition between the processes of fission and particle evaporation within
the framework of the statistical model. This depends on the particular calculating
algorithm used. For the decay of a heavy highly fissioning compound nucleus, the
assumption that the main process competitive to the fission is neutron evaporation
will be a good approximation to calculate the value $W_{\rm sur}$.
In this case, there exists a simple phenomenological method for calculating
the survival factor $W_{\rm sur}$, which can be
written as the product of the ratios $
\left\langle {\Gamma _n /\Gamma _{tot} } \right\rangle  =
\left\langle {\Gamma _n /(\Gamma _n  + \Gamma _f } \right\rangle
\approx \left\langle {\Gamma _n /\Gamma _f } \right\rangle
$ averaged over the
evaporation cascade. The magnitude of the fissionability averaged over the evaporation
cascade
\begin{equation}
\label{eq:form10}
\left\langle {\Gamma _n /\Gamma _f } \right\rangle  = \left[
{\mathop \Pi \limits_{i = 1}^x \left( {\Gamma _n /\Gamma _f }
\right)} \right]^{1/x} ,
\end{equation}

is derived from experimental cross-sections $\sigma _{xn}$ with the help of equation (9).
It depends weakly on the values $E^*$ and $x$. Then the survivability of the heavy nucleus
in terms of that approach can be written in the following analytical form:

\begin{equation}
\label{eq:form11}
W_{\rm sur} (E^* ,l) \approx P_{xn} (E^* ,l) \cdot \prod\limits_{k =
1}^x {\left\langle {\frac{{\Gamma _n (E_k^* )}}{{\Gamma _f (E_k^*
)}}} \right\rangle _k ,}
\end{equation}

where $x$ is the number of evaporated neutrons; $P_{xn}$ -- the probability of evaporating
$x$ neutrons;
$k$ -- the index of the evaporation cascade stage.

In work [33], a large body of experimental data on few-neutron emission reactions
was analysed, and semi-empirical systematics for the value
$\left\langle {\Gamma _n /\Gamma _f } \right\rangle $
averaged over
the cascade were obtained, which permitted describing the experimental data known
by that time with an accuracy of up to the factor $2 \div 3$. The value 
$P_{xn}$ in Eq.(\ref{eq:form11}) was
calculated following [34].

There exist other approaches to calculating the survivability of excited compound
nuclei in the de-excitation process within the framework of the statistical model.
Brief mention can be made of the latest calculations of that value for the considered
compound nucleus of $^{256}$No in works [35, 36]. In work [35], into the formula for
calculating survivability
\begin{equation}
\label{eq:form12}
W_{\rm sur} (E^* ) = \prod\limits_{k = 1}^x {\left( {\frac{{\Gamma _n
(E_k^* )}}{{\Gamma _{tot} (E_k^* )}}} \right)_k  \cdot P_x (E^* )}
\end{equation}

the standard expressions were substituted from the statistical model for the
probability of evaporating neutrons, protons, $\alpha$-particles,
$\gamma$-quanta and fission.
A system of integrals inserted into one another (due to the number of stages in
the evaporation cascade) was arrived at. Then numerical integration was carried
out to calculate the magnitude of the survivability. Details can be found in the
cited paper. In work [36], after a number of reasonable simplifications and integration,
simple expressions were derived for the probability of evaporating neutrons and fission.
The factor $P_{xn}$ in formula~(\ref{eq:form12}) was calculated on the basis of Monte-Carlo method,
simulating the probability of emitting neutrons of given energy. The neutron
distribution over the energy $e$ carried out of the excited compound nucleus was
taken in both cited papers to be the Maxwellian distribution:
\begin{equation}
\label{eq:form13}
P_n (E^* ,e) = C\sqrt e \exp \left( { - \frac{e}{{T(E^* )}}}\right)
\end{equation}

In our case, we used the calculations of the  $W_{\rm sur}$ value within the framework
of a standard statistical model, which takes into account evaporation of not only
neutrons but also of charged particles and $\gamma$-quanta. Since we study the decay of
compound nuclei of considerable angular momentum, we made use of a quasi-classical
version of the statistical model [37]. In that version, the probability of emitting
particle $\nu$(p, $\alpha$, d, t, $^3$He) of energy $e_\nu$ and angular
momentum $\vec l$ in a unit time from
a compound nucleus of angular momentum $\vec L$ and excitation energy $E^*$ in the direction $\vec n$
is written as follows:
\begin{equation}
\label{eq:form14}
P_\nu  (\vec l,\vec n,e_\nu  ) = (2s_\nu   + 1)\frac{{\mu _\nu  e_\nu  }}
{{\pi ^2 \hbar ^3 }}\frac{{\rho _d (E_d^* ,L_d )}} {{\rho _m
(E_m^* ,L_m )}} \cdot d\sigma _{\rm inv} (\vec n,e_\nu  )
\end{equation}

where $\mu$ is the reduced mass of the system; $s$ -- the spin of the emitted particle; the
symbols $m$ and $d$ indicate the mother and daughter nuclei, correspondingly.

Integrating the latter expression over the energy of the emitted particle gives
the total probability and, thus, the partial width for emitting particles from the
compound nucleus ($\Gamma_{n,\nu} = P_{n,\nu}/\hbar$). To calculate the partial width for emitting neutrons,
charged particles ($\nu =$  p, $\alpha$, d, t, $^{3}$He),
$\gamma$-quanta and fission, the following
expressions were used (see, for example, in reference [38]):
\begin{eqnarray}
\label{eq:form15}
\Gamma _n (E^* ,L) \approx \frac{{(2s + 1)\mu }} {{(\pi \hbar )^2
\rho _m (U)}}\int\limits_0^{U - B_n } {\sigma _{\rm inv} (e_n )\rho
_d (U - B_n  - e_n )e_n de_n ,} \nonumber \\
\Gamma _\nu  (E^* ,L) = \frac{{(2s_\nu   + 1) \cdot \mu _\nu  }}
{{\pi ^2 \hbar ^3 \rho _m (U)}}\int\limits_{v_\nu  }^{U - B_\nu
} {\sigma _{\rm inv} (e_\nu  ) \cdot \rho _d (U - B_\nu   - e_\nu  )
\cdot e_\nu  de_\nu  } \\
\Gamma _\gamma ^{} (E^* ,L) = \frac{3} {{(\pi \hbar c)^3 \rho _m
(U)}}\int\limits_0^U {\sigma _{\rm inv} (e_\gamma  ) \cdot \rho _d (U
- e_\gamma  ) \cdot e_\gamma ^2 de_\gamma  } \nonumber \\
\Gamma _f (E^* ,L) \approx (2\pi \rho _m (U))^{ - 1}
\int\limits_0^{U_s  - B_f } {\rho _s (U_s  - B_f  - \varepsilon)d\varepsilon ,} \nonumber
\end{eqnarray}

where $U$ is the heat energy of the mother nucleus. The cross-section of the reverse
reaction $\sigma _{\rm inv}$, (reaction of the absorption of
particles by the nucleus) was taken
following work [39]. While calculating the partial width for $\gamma$-quantum emission,
we only took account of emission of dipole $\gamma$-quanta. In this case, the inverse 
cross-section was taken in the form of a giant dipole resonance [40]. In the expression for
the partial width of fission, the heat energy $U_s$ and rotation energy $E_r^s$ at the saddle
point are related to each other by the relation $U_s=E^*- E_r^s$.
While calculating $\Gamma _f$,
account was taken of the fact that the rotation of the nucleus causes the fission
barrier decrease: $B_f(L)= B_f(0)-(E_r-E_r^s)$. To calculate the level density, the known
Fermi-gas expression (see, for example, [41]), was used
\begin{equation}
\label{eq:form16}
\rho (E^* ) = \frac{{\sqrt \pi  }} {{12a^{1/4} (E^*  - \delta
)^{5/4} }}\exp [S(E^* )],
\end{equation}

In Eq.~(\ref{eq:form16}), the nucleus entropy
$S$ at the excitation energy $E^*$ is taken to be described
by the relationship $S = 2aT$, using the relation between the temperature and excitation
energy of the nucleus $E^* = aT^2$; $\delta$ is the even-odd correction [41].
The level density
parameter $a=\pi^2g_0/6$ is expressed as a function of the one-particle density near Fermi's
energy $g_0 = f(E_f) = const$. That the influence of shell effects
in level density decreases as excitation energy increases was taken account
of according to the phenomenological expression from [41]:
\begin{equation}
\label{eq:form17}
a(E^* ) = \tilde a[1 + f(E^* )\Delta W/E^* ],
\end{equation}

Here $f(E^*)=1-exp(-\gamma \cdot E^*)$, $\Delta W$ is the shell correction to the
nucleus mass; $\tilde a$ is the
asymptotical magnitude of the level density parameter within the Fermi-gas model,
which can be written in the general form as
$\tilde a =A\cdot(\alpha+\beta A^n)$. Empirical values for the
parameters $\alpha = 0.148~$MeV$^{-1}$; $\beta = -1.39\cdot 10^{-4}$~MeV$^{-1}$;
$\gamma = 0.061$~MeV$^{-1}$ at $n=1$ were found in work [42] from analysis of
data on the level density with regard to the contribution
of the collective states to the total level density.

To calculate $W_{\rm sur}$, the fission barrier for compound nuclei should
be known. The fission barrier $B_f$ is calculated as the sum of the
liquid-drop component and the shell correction $\Delta W: B_f = B_f^{LD} + \Delta W$.
For the compound nuclei of the heavy transfermium region, the
liquid-drop component is a very small quantity.

In order to take into account the dependence of the shell correction on the
excitation energy, we used  the following formula:
$B_f(E^*) = B_f^{LD} + \Delta W\cdot exp(-\gamma E^*)$.
In our calculation, we used the value $B_f(E^*=0)$ from work [43].

\subsection{    Scheme of calculating the de-excitation process for an excited
compound nucleus on the basis of the Monte-Carlo method}

In this work, the de-excitation process for a nucleus was calculated by applying an
approach based on simulating the random value --- the Monte-Carlo method. Such an approach
was successfully used to calculate the decay of heavy nuclei in [44]. In our opinion, it
reflects the random, statistical nature of particle evaporation or fission in the most
adequate way.

The angular momenta of the compound nuclei produced in a complete fusion reaction will
be distributed over the value $L$, the vector $\vec L$ lying in the plane perpendicular to the
ion beam. With the help of those random numbers, the value of the angular momentum and
its orientation in space were simulated. Then the maximum residual energies in the
different channels of the statistical decay were simulated
\begin{equation}
\label{eq:form18}
E_\nu ^{(\max )*}  = E^*  - E_{\rm rot}  - B_\nu   - V; \quad
E_f^{(\max )*}  = E^*  - E_{\rm rot}  - B_f
\end{equation}

For all the $E_\nu^*>0$, the type of emitted particle was also randomly 
chosen --  $\nu = n, p, d, ^3$He, $\alpha$-particle  or
$\gamma$-quantum --- on the basis of formulae~(\ref{eq:form15}). After the type 
of de-excitation process
(the decay channel) was found, and if no fission occurred, the characteristics of
the evaporated particles or $\gamma$-quanta were simulated. So their kinetic energy,
carried-away orbital momentum and emittance angle $\theta$ were found.
For a given type of particle,
the magnitudes of $e_\nu,  \vec l$ and $\cos \theta$ were simultaneously sampled
using three random
numbers, and then rejection was performed with the help of a fourth random number
according to the three-dimensional probability density given by the expression [45]

\begin{equation}
\label{eq:form19}
W(e_\nu ,l_\nu ,\cos \theta ) \sim l \cdot \exp \left[
{2\sqrt {a(E^* - E_{\nu} - (L^2 + l^2 )/2J + Ll \cos \theta /J)}}\right]
\end{equation}

In a coordinate frame with a $Z$ axis parallel to $\vec L$, the azimuthal angle of the vector
$\vec l$ was simulated. The azimuthal angle of the escaped particle was simulated in a coordinate
frame with a $Z$ axis parallel to $\vec l$. The fission process was
accounted for with the help of
the weight  functions
$FU = \prod\limits_{\nu  = 1}^x {\left[ {1 - \Gamma _f /\Gamma _{tot} } \right]} $,
which is especially convenient for highly
fissioning nuclei and significantly saves computing time. All the values so found were
brought to the centre-of-mass system of the colliding nuclei, and then the characteristics
of the residual nucleus were calculated
 
$\vec L_d=\vec L_m-\vec l;\quad E_d=E_m-B_n-e_n-(L^2_d-l^2)/2J;\quad A_d=A_m-A_n;\quad Z_d=Z_m-Z_n$

Then for that residual nucleus, the maximum residual energies were calculated for all
the decay channels: particle emission, $\gamma$-quantum emission and fission. Among all the
processes energetically allowed, the de-excitation of the nucleus was again simulated
up to $E_s^*>0$.

Presented in Fig.~\ref{fig:fig12} are the calculated results for the 
probability of neutron evaporation
from a nucleus of $^{256}$No as a function of excitation energy.
Also shown in the picture are the experimental values derived from the neutron evaporation
cross-sections by the so-called pair reaction method (for details on the pair reaction
method, see [46], from which that figure is borrowed). It is seen from the data presented
in the figure, that for the larger excitation energy of the nucleus, it is necessary to
take account of the shell correction being dependent on $E^*$. For $E^*>40$~MeV,
shell effects in such a heavy nucleus as $^{256}$No damp out almost completely.

Shown in Fig.~\ref{fig:fig13} are the calculated results for the survivability of
an excited compound nucleus of $^{256}$No obtained within the framework of the
statistical model with three different
approaches. The solid curve is the calculated result from this work. The dashed curve is
the calculated result from work [35]. The dotted curve is borrowed from work [47], the
model used being described in work [36]. It is seen from the figure that the different
calculating methods give close magnitudes for fissionability in the region of excitation
energies from $13$ to $35$~MeV, but as energy increases the magnitudes from the dotted curve
begin to exceed the magnitudes from the other curves. This might be connected with the
fact that the approach elaborated in [36] only takes account of the fission and
neutron evaporation channels, whereas at the energies higher than $35$~MeV, the channels with
evaporation of charged particles begin to manifest themselves.

Summing up, in Fig.~\ref{fig:fig5} we present a comparison both of the experimental
capture cross-sections and of the excitation functions for evaporation of
different numbers of neutrons measured in work [20] with our theoretical
calculations. It is clearly visible that our
approach adequately describes the measured data available for the reaction
$^{48}$Ca $+^{208}$Pb$ \rightarrow ^{256-xn}$No both on fusion and on the excitation function
with particle evaporation.

\section{ Analysis and discussion}

\subsection{ MED of fragments of the heated nucleus}

In section \ref{sec:res}, we have shown that the MEDs of fragments are
complex in structure and
introduced the term {\em ''quasi-fission shoulders''}. The introduction of that term was not
incidental, and not only due to the fact that their characteristics differ from
those of the classical fission. There was another reason for that.
 In works [48-50], the reactions $^{40}$Ar $+^{208}$Pb and
$^{50}$Ti $+^{208}$Pb, which are close
to our case, but involve lighter and heavier ions, were studied. It was shown that
the angular distributions of fragments (ADF) for high mass/charge asymmetries,
which  exactly correspond to the ''shoulders'', have a pronounced forward-backward
 asymmetry. It indicates that the process is a non-equilibrium one
 [7, 8, 48-50]. At the same time, the ADFs for fragments of the mass/charge
 lying in the vicinity of $A_{\rm CN}/2  (Z_{\rm CN}/2)$ are symmetric
about $90^\circ$. So, in our case the ''shoulders'' can be believed with good reason to be also
of a quasi-fission nature.

In work [51], it was observed that the picture for the ADF in the reaction
$^{40}$Ar + $^{197}$Au is similar to that in the reaction $^{40}$Ar $+^{208}$Pb
and shown that
in both cases the ADF is representable as the sum of two components:
a compound one and a quasi-fission one. To assess the partial cross-sections
for reaction product yields, in works [52, 53] a similar procedure was also
applied to the mass distributions of fragments in the reactions
$^{208}$Pb$ + ^{50}$Ti [5], $^{51}$V$ + ^{197}$Au [52],
and $^{197}$Au$ + ^{63}$Cu [53].
In recent work [54], it was also shown for the reaction
$^{48}$Ca$+^{168}$Er that experimentally observed MEDs of fragments
are the sum of classical fission and quasi-fission components.
In present work  we will follow this way while analysing the MEDs of
fragments as well  and show its validity.

\subsubsection{Symmetric fission properties}

For the nuclei of the second half of the periodic table, the properties of classical
symmetric fission, characteristic of heated nuclei, are quite well studied for
reactions with heavy ions and known from numerous both experimental
[5, 7, 8, 18, 19, 23, 24, 55-63] and theoretical [21, 22, 64-68] works.
Their MD of fragments is a one-humped gaussian-like curve, whose variance, $\sigma^2_{M}$,
increases approximately in proportion to the increasing temperature of a fissioning
nucleus: $\sigma^2_{M} \sim T$ and also $\sigma^2_{\rm TKE} \sim T$,
whereas $\langle TKE \rangle$ is practically independent of
$E^*$ (or $T$). Experiments showed that an increase in the angular momentum $l$ of a
fissioning nucleus results in an additional increase in the $\sigma^2_{M}$ for heavy nuclei
with $Z^2/A > 30$ [18, 55, 58, 63] and in a decrease in the $\sigma^2_{M}$ for light nuclei
with $Z^2/A < 30$ [59, 61, 68]. In review [24], the properties of the MEDs of fragments
for heated nuclei with $A_{\rm CN}> 100$ were analysed and systematised. In works [18, 24, 63],
in studying the fission of the superheavy compound nucleus of $^{260}$Rf ($Z=104$) produced
in various reactions: $^{16}$O + $^{244}$Cm,
$^{20}$Ne$ + ^{240}$Pu, $^{24}$Mg$ + ^{236}$U it was found that the
properties the MEDs of fragments --- a rise in $\sigma^2_{M}$ and $\sigma^2_{\rm TKE}$with the
increasing $T$ and $l$, the
independence of $\langle TKE \rangle$ of those variables - fit in with the
known picture of classical
fission.

To find the characteristics of MEDs for the symmetric fission of $^{256}$No
(Figs.~\ref{fig:fig7} and \ref{fig:fig8}), we approximated the MD of fragments
with a gaussian $G$.  On the left-hand side of Fig.~\ref{fig:fig14}, the mass
distributions of fragments, depicted on a logarithmic scale,
are given in relation to the parameter $(M-A/2)^2$. Being so represented, the gaussian
is a linear function. In Fig.~\ref{fig:fig14} it is quite clearly visible that at the parameter
value $(M - 128)^2 \cong 800-900  (M_L \cong 100)$, the experimental points start to deviate
upwards from $G$, which adequately approximates the beginning part of the MD. In
the central panels of Fig.~\ref{fig:fig14}, the $G$ approximation for the MD is shown on a linear scale.
In Table~\ref{tab:table1}, given are the values of the variances of the gaussian distributions
($\sigma^2_{M}$) for five magnitudes of $^{48}$Ca ion energy. On the right-hand side
of Fig.~\ref{fig:fig14}, shown are
(only for light fragments) the ''quasi-fission shoulders'' derived in a ''pure'' state by
subtracting the $G$-function from the experimental yield of fragments:

\begin{equation}
\label{eq:form20}
Y_{\rm QF}(M) = Y_{\rm exp}(M) - Y_{\rm G}(M).
\end{equation}

Let us consider the properties of the MED for symmetric fission, or more precisely,
the experimental dependences of $(\sigma^2_{M})_G, \langle TKE \rangle, \sigma^2_{\rm TKE}$
on $T$ and $l$. In Fig.~\ref{fig:fig15}, the
dependence is shown of those characteristics on the nucleus temperature at the
scission point $T_{\rm sc}$, which we find, as in a great number of works (for example
[24, 63,69] from

\begin{equation}
\label{eq:form21}
T_{\rm sc} = (E_{\rm sc}/\tilde a)^{1/2}, \quad
E_{\rm sc} = E^*_{LD} + Q_{f}-{\langle TKE\rangle}-
{\langle \nu _{\rm pre}\rangle}{\langle E_{\nu}\rangle} - E^{\rm sc}_{\rm rot}-
E_{\rm def}+E_{\rm diss} ,
\end{equation}

\begin{equation}
\label{eq:form22}
  E^*_{\rm LD} = E_{\rm CM} + (\Delta M_i + \Delta M_t -\Delta M^{\rm LD}_{\rm CN}),
\end{equation}

where $E^*_{\rm LD}$ is the nucleus excitation energy measured from the ground state of the
liquid-drop model ; $\Delta M_i$ and $ \Delta M_t$ the experimental magnitudes of the mass
defects for the nuclei; $\Delta M^{\rm LD}_{\rm CN}$ the liquid-drop magnitude of the mass defect
for the compound nucleus of $^{256}$No from work [70]. $\tilde a=0.093A_{\rm CN}$ is the level density
parameter [41,71]. $Q_f$ is the reaction energy for the symmetric scission of the nucleus
into fragments. ${\langle TKE \rangle}$ is the measured average magnitude of the $TKE$.
${\langle\nu _{pre}\rangle}{\langle E_\nu \rangle}$ is the
average energy carried away by the pre-fission neutrons calculated as proposed in [15].
The value of ${\langle\nu _{pre}\rangle}$ is found from systematics [24] and given in Table~\ref{tab:table1}
for all the magnitudes of $^{48}$Ca-ions energy. $E^{\rm sc}_{\rm rot}$ is the nucleus
rotation energy at the scission point. $E_{\rm def}$ is the energy spent on
deforming the fragments. $E_{\rm diss}$ is
the energy dissipated from the collective degrees of freedom into one-particle
degrees of freedom.

Let us explain some components in expression Eq.~(21). The nucleus rotation energy at
the scission point was found from expression from Ref.~[55]
\begin{equation}
\label{eq:form23}
E^{\rm sc}_{\rm rot}  = l^2 \hbar^2/ 2J^{\rm sc}_{\perp} + T_{\rm sc}/2,
\end{equation}

where $J^{\rm sc}_{\perp}$ is the nucleus moment of inertia relative to the axis
perpendicular to the fission axis. In the end, the matter comes to finding
$J^{\rm sc}_{\perp}$.
The LDM by Strutinski [72] defines a location where scission occurs as a
critical point associated with a loss of stability against breakup into
two fragments and uses $J_0$ units for $J^{\rm sc}_{\perp}$
($J_0$ is the rigid body moment of inertia of the spherical nucleus). For the nucleus of
$^{256}$No, $J^{\rm sc}_{\perp}\cong 4.3 J_0$. Hence it follows that
\begin{equation}
\label{eq:form24}
E^{\rm sc}_{\rm rot}\cong E^0_{\rm rot}/4,3 + T_{\rm sc}/2.
\end{equation}

From the classical rotating liquid-drop model [73], it is known that

\begin{equation}
\label{eq:form25}
E^0_{\rm rot}= 34,540 (l^2 / A^{5/3})
\end{equation}

It should be noted that in the framework of LDM [72], the scission
point has by no means non-zero neck thickness. Scission occurs when the
neck is thick enough, the common elongation not changing. So scission
does not result from the neck gradually thinning with the increasing
deformation of a fissioning nucleus, as is often assumed in a lot of works.

 That is, a breakup is by no means preceded by a gradual decrease in
the neck thickness accompanied by an increase in the deformation of the fissioning
nucleus, contrary to what is frequently assumed in a
lot of works. Such theoretical definition of
the scission point is supported by the analysis of various characteristics
of measured energy distributions of fragments when taking into account
viscosity  of nuclear matter [24, 74-76].

As for the parameters $E_{\rm def}$ and $E_{\rm diss}$ in expression (21),
it can be said that there
is currently no general agreement on those values --- they are dependent on the model
used, the nature and magnitude of nuclear viscosity (two-body or one-body viscosity).
But account should be taken of the fact that $E_{\rm def}$ and $E_{\rm diss}$
compensate each other to
some extent. Therefore, following works [24, 63, 77],
we took $E_{\rm def} = 0$ and $E_{\rm diss} = 0$.
The temperature $T_{\rm sc}$ found according to Eq.~(21) is given in Table~\ref{tab:table1}. 
It should be noted
that $T_{\rm sc}$ rises very slowly in those cases at the high energy of bombarding ions when
$\nu_{\rm pre}$ are evaporated. That property of fissioning heavy nuclei, noticed in works
[18, 24, 77], results in the fact that reaching the scission point, the nucleus
heats up only slightly in spite of the increasing energy of the ions.

In Fig.~\ref{fig:fig15}(a), it is clearly visible that within the limits of experimental error,
the magnitudes of $\langle TKE \rangle$ are independent of $T_{\rm sc}$ (excitation), which is characteristic
of the fission of heavy and superheavy nuclei. For $T_{\rm sc} \geq 1.83$~MeV, the varience
$\sigma^2_{\rm TKE}$ (Fig.~\ref{fig:fig15}(b) increases proportionally to $T_{\rm sc}$.
The shaded magnitudes of $\sigma^2_{\rm TKE}$
belong in the region peculiar to classical fission. That region was determined in
works [18, 23, 24, 63] for the fission of the heavy nuclei produced in reactions with
"light" heavy ions ($A_i < 26$) with taking into account an increase of $\sigma^2_{\rm TKE}$ with $l$.
 For the temperature $T_{\rm sc} \leq 1.82$,
there is no decrease in the magnitude of $\sigma^2_{\rm TKE}$, and at $T_{\rm sc} = 1.75$
($E_{\rm  lab} = 211 $~MeV), whereas the magnitude
of $\sigma^2_{\rm TKE}$ even grows. As has been noted in section~\ref{sec:res}, this points
to the influence of shell effects on the MED of fragments for the sufficiently low-energy
fission of $^{256}$No.

In Fig.~\ref{fig:fig15}(c), the variance of fragment masses $(\sigma^2_{M})_G$ is shown as a
function of $T_{\rm sc}$ for
the five largest magnitudes of the energy of $^{48}$Ca ions (Figs. \ref{fig:fig6},
\ref{fig:fig7}, \ref{fig:fig14}).
The cross-hatched region of the ($\sigma^2_{M})_G$ magnitudes, as in Fig.~\ref{fig:fig15}(b),
belong in the region typical of normal
fission. Since $T_{\rm sc}$ increases only slightly with increasing ion energy, then, as shown
in works [18, 24, 63], the increase in ($\sigma^2_{M})_G$ is to a large degree connected with the
increase in the angular momentum $l$ of the fissioning nucleus, and
($\sigma^2_{M})_G \sim \langle l^2 \rangle$.
Therefore for the temperature $T_{\rm sc} = 1.93$~MeV ($E_{\rm  lab} = 242$~MeV),
the variance ($\sigma^2_{M})_G$
shows a sharp enough rise. That increase in variance lies within the boundaries of
what can be expected for normal fission.

So we come to the conclusion that the characteristics of the MEDs of fragments for the
fission of $^{256}$No for the region of symmetric masses fit in
with ordinary classical fission.

\subsubsection{Properties of the quasi-fission process}

In contrast to the properties of classical fusion-fission, the characteristics of
quasi-fission process depending on the nucleon composition, excitation energy and entrance
channel energy have been poorly studied. First of all, it is true for the energy
characteristics of QF.
Early it was believed, that $TKE$ for the QF process did not differ
from that for the FF process. However, on analysing a great number of reactions induced by various
ions, it has been shown recently in work [24] that for QF, the $TKE$ of fragments is on
average larger. {\em Experimental} dependences  $TKE_{\rm QF}(M)$ and
$\sigma^2_{\rm TKE(QF)}(M)$ were not practically considered.
 Just quite recently, it has been shown in work [54] that in the reaction
$^{48}$Ca$+^{168}$Er$ \to ^{216}$Ra at $E^*=40$~MeV, about $30 \%$ of
the fragments are associated with quasi-fission and have $TKE$ higher than
the $TKE$ of the fragments produced in a classical FF process and that this
experimental fact, by analogy with low-energy fission, can be
accounted for by QF being strongly influenced by shell effects. Besides,
it has been demonstrated in [54] that the properties of the MED of
fragments for $^{216}$Ra formally satisfy the standard hypothesis about independent fission
modes [55, 78, 79]; in that case, there are just QF and
normal FF. Our version of the DNS model corresponds to this concept:
FF and QF are assumed to be competitive processes and the MED of
fragments for a process is considered not to be influenced by the other
process.

\begin{figure}[p]
\includegraphics[width=6cm]{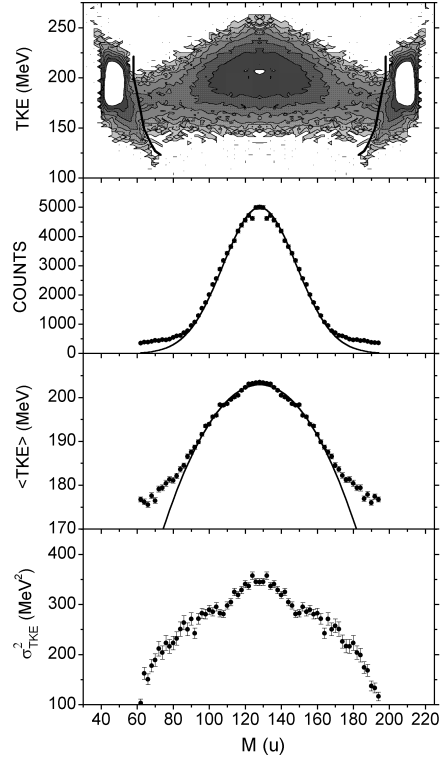}
\caption{\label{fig:fig7} The same as in Fig.6, but for the energy of $^{48}$Ca-ions
$E_{\rm lab}=230$~MeV. The fragment-like events 
in the $M-TKE$ matrix are enclosed by solid curves. The characteristics of those
fragments are given below.
}
\end{figure}

\begin{figure}[p]
\includegraphics[angle=-90,width=10cm]{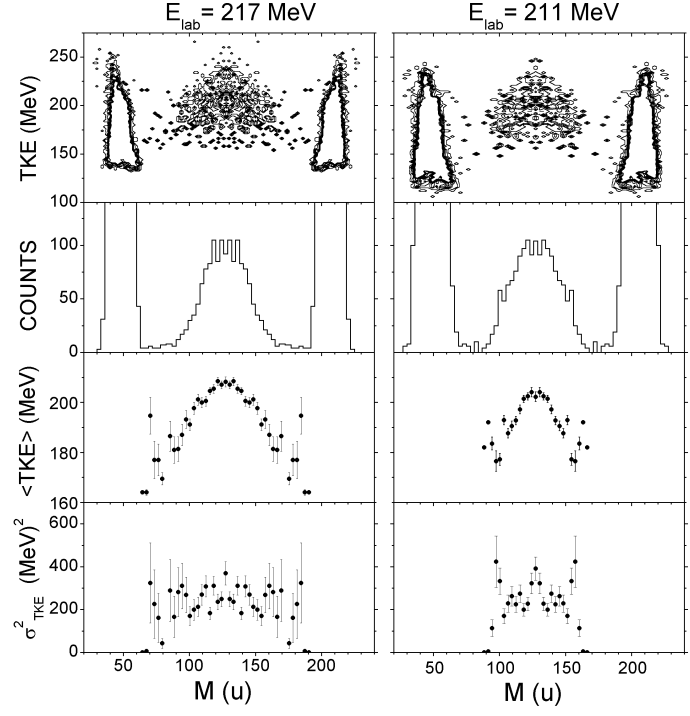}
\caption{\label{fig:fig8} The same as in Fig. 6, but for the two lowest energies
of $^{48}$Ca-ions.}
\end{figure}

\begin{figure}[p]
\includegraphics[width=7cm]{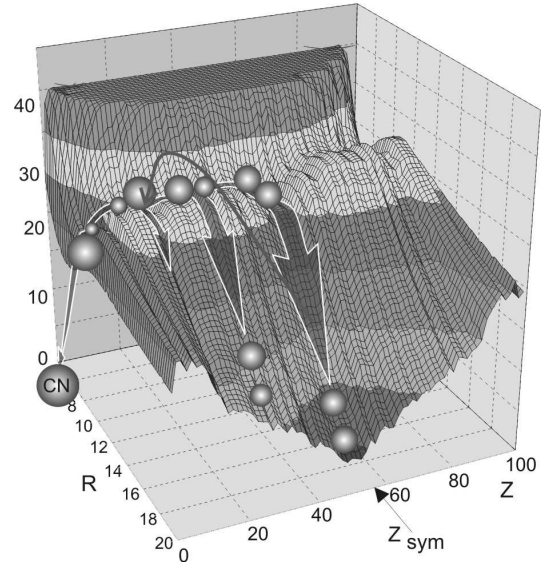}
\caption{\label{fig:fig9} PROXIMITY potential as a function of the charge number $Z$ and
the centre-to-centre distance of fragments. The input channel and
the ways for DNS to evolve are indicated with arrows.}
\end{figure}

\begin{figure}[p]
\includegraphics[width=10cm]{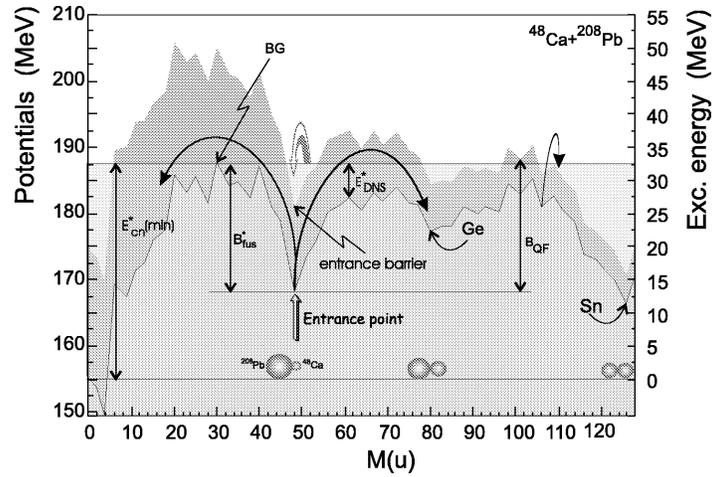}
\caption{\label{fig:fig10} By way of illustration, the driving potential is shown (lower
curve) of the dinuclear system for the reaction $^{48}$Ca$+^{208}$Pb as a
function of fragment mass M.}
\end{figure}

From Fig.~\ref{fig:fig7}, it is seen that the fragment masses in the region of the 
quasi-fission shoulders have $TKE$ higher than that expected for a FF process, which
agrees with work [54]. Therefore in this case the MED is also
decomposable into components, and this can be done in accordance with
Eq.~(\ref{eq:form20}) and an expression from [54]:
\begin{equation}
\label{eq:form26}
TKE_{\rm QF}=[TKE_{\rm  exp}-TKE_{\rm FF}(Y_{\rm FF}/Y_{\rm exp})]/(Y_{\rm QF}/Y_{\rm exp})
\end{equation}
Here everywhere $M$-dependence is meant.

The decomposition was done for the energy $E_{\rm lab}=230$~MeV, for
which the data were statistically sufficient to the largest degree. The
experimental dependence $\langle TKE \rangle (M)$ in Fig.~\ref{fig:fig7} was fitted with a parabolic
curve [55].
\begin{equation}
\label{eq:form27}
TKE(M)=TKE(A/2)(1-\eta^2)(1+\rho \eta^2),\quad \eta=(M-A/2)/(A/2),
\end{equation}

where $\rho$ is an empirical parameter that defines the extent to which
the width of this parabolic curve deviates from the predictions of LDM
[21], for which $\rho=0$. In our case $\rho=0.25$, which is in good
agreement with the results obtained for classical fission [55].

 The mass distribution of light QF fragments is presented at the right of
Fig.~\ref{fig:fig14} for four values of energy. Fig.~\ref{fig:fig16} shows the decomposed MED of
light fragments for the energy $E_{\rm lab}=230$~MeV. Shown in Figs.~\ref{fig:fig16}(a)
 and \ref{fig:fig16}(b) are the mass distribution for QF and the dependence 
$\langle TKE \rangle (M)$ respectively. 
It is clearly seen that the $TKE_{\rm QF}(M)$ is higher by $7-
15$~MeV than $TKE_{\rm FF}(M)$ and that on the whole it is also parabolic in
form, which is also a formal similarity to the modes of classical low
energy fission [80]. However the width of the  parabolic dependence $TKE_{\rm QF}(M)$ is
greater than $TKE_{\rm FF}(M)$ (parameter $\rho=0.6$).

It would appear natural to hypothesize, as in [54], that the properties of
the MED for QF can be accounted for by shell effects, which is also
predicted by our version of the DNS model. In Fig.~\ref{fig:fig16}(a), the mass number
 of the closed spherical shells $Z=28, N=50$ are indicated
with arrows. They were calculated on the basis of the simple assumption
that charge is proportional to mass (unchanged charge density). In this
case, only a light ''quasi-fission shoulder'' is associated with closed shells. The
complementary heavy fragment belongs to the mass region $M \sim 160-190$
and is non-spherical.  The influence of shells with $Z=50$ and
$N=82$ near the mass $M=A/2$ will be considered in the next section.

The properties of QF similar to those discussed, with a maximum yield of
fragments at $Z_{L(QF)}\sim 28$ and $M_{L(QF)} \sim 75-85$, have already been
revealed in the reactions $^{51}$V$+^{197}$Au and $^{197}$Au$+^{63}$Cu
in works [52, 53] respectively. The effects under discussion seem to
manifest themselves in the mass region of quasi-fission fragments $M \sim 70-85$ in
the reactions $^{58}$Ni$+^{124}$Sn [81],
$^{124,136}$Xe$+^{58,64}$Ni [82], $^{208}$Pb$+^{50}$Ti,$^{64}$Ni
[5], $^{40}$Ar$+^{208}$Pb [83] as well as at the forward angles in the
reaction $^{64}$Ni$+^{197}$Au [84]. There arises an interesting
situation with the $^{48}$Ca$+^{168}$Er reaction [54]. In this case, the
complementary heavy fragment, its mass $M_H(QF)\approx 130-140$, is close
to the magic spherical shells with $Z=50$, $N=82$ in nucleon
composition, therefore, unlike the case discussed, both the light fragment
and the heavy fragment are represented in the quasi-fission shoulders, which
results in more pronounced QF. The same is probably true for the
$^{164}$Ho$+^{58}$Ni reaction [85], characterized by maxima just in
the same fragment mass regions at the forward and backward angles
peculiar to QF. At the same time, at the angles close to $90^\circ$ in the
vicinity of the mass $M \sim A/2$, as in our measurements, a symmetric
near-Gaussian component, which we as well as the authors of [55, 85]
believe to be associated with FF, is clearly visible.

From this point of view, the reaction $^{40}$Ar$+^{208}$Pb
$\rightarrow ^{248}$Fm at $E_{\rm lab}=247$, discussed in work [83], is very
interesting. This reaction is close to the reaction under consideration, and
its MDs show clearly defined quasi-fission shoulders. This feature has
somewhat escaped attention before. However in our opinion the results of
work [83] are of outstanding importance in understanding the properties
of QF. In this work, the MDs of fission fragments for $^{248}$Fm were
measured for two angles: $-24^\circ$ and $-46^\circ$. Having analyzed those data,
we extracted the QF component from the total yield of fragments. The
results obtained are presented in Fig.~\ref{fig:fig17}, at the left of which the MD of
fragments is shown for forward angle registration. The quasi-fission shoulder
is perfectly well visible. That this shoulder is associated with quasi-fission
rather than with an exotic classical fission mode, follows from work [48],
already cited. In this work, it was shown that the angular distributions of
the fragments from the shoulders are highly asymmetric; whereas in the region
of symmetric charges/masses, the angular fragment distributions fall into
the pattern of normal fission. The central symmetric peak in Fig.~\ref{fig:fig17},
associated with a FF process, can be approximated, as in the case under
discussion, by a Gaussian to a good accuracy. The extracted QF
component, indicated with open circles, does not merge with the 
quasi-elastic peak as in the case of $^{48}$Ca-projectiles and is distinctly
limited on both the light and heavy fragment mass sides; and again the
quasi-fission shoulder is composed of fragments with near-magic numbers
close to $Z \approx 28$ and $N \approx 50$. At the right of Fig.~\ref{fig:fig17}, the MD of
fragments is shown for the same nucleus but for a larger angle of
measurement. In this case, the quasi-fission shoulders are less prominent but
the lower portion of the MD is clearly seen to be wider, in the same way
as in the case under discussion at the energy of $^{48}$Ca ions
$E_{\rm lab} = 242$~MeV. The QF is characterized by the same mass range of
light fragments as that shown at the left of Fig.~\ref{fig:fig17}. Unfortunately, work
[83] does not discuss the energy distributions for this particular reaction.
It is noteworthy that the $^{248}$Fm compound nucleus has the initial
excitation energy $E^* \cong 78$~MeV and that the number of pre-fission
neutrons in the quasi-fission shoulders region is only $1.2$ [83]. Thus,
although $E^*$ is high enough, the conclusion can be drawn that shell
effects play a greater role in QF than in classical FF, which we have
already shown in work [54] for $^{216}$Ra.

It should be recalled that although the role of shell effects in QF has been
discussed for a long time [6, 86-88], the interaction of massive ions with
actinide targets (or U ions in reactions of inverse kinematics) is basically
studied. The MDs of fragments of such reactions are characterized by the
presence of doubly magic lead, $M_H \sim 208$ ($Z_H \sim 82, N_H \sim 126$),
which is variously accounted for. For example, the authors of works [6,
87], studying the reactions $^{40}$Ar, $^{48}$Ca$ + ^{238}$U, reasoned
that the peak associated with lead is due to the further sequential fission
of fragments with mass $M_H = 215-230$; whereas the authors of work
[88] hold to the idea that this peak owes its existence to strong lead shells
nevertheless. To return in our case to the reaction $^{48}$Ca $+^{208}$Pb at
$E^*=33$~MeV, it is unlikely for fragments of any mass number to
undergo sequential fission [41, 71]. On the contrary, increased yield of
fragments with mass $M = 60-80$ is observed, their $TKE$ being higher
compared to the $TKE$ expected for a FF process.

Recently, studying the reactions $^{48}$Ca$ + ^{238}$U, $^{244}$Pu,
$^{248}$Cm at $E^*\sim 30$~MeV [16, 89, 90], we have found corroboration
of the fact that the properties of the QF of superheavy nuclei are
influenced by magic shells, both the shell structure of heavy fragments
(nuclei close to $^{208}$Pb) and light fragments (nuclei close to $^{78}$Ni)
being of importance, which is shown in [54] as well. So, based on the
results of the experiment discussed and previous research, we can
conclude that in QF, closed spherical shells play an important role and
that shell effects damp out much differently than in known and usual
classical fission. The questions immediately arise ''Why so?'' and ''Why
is all this possible?''.

We believe that the answers, at least qualitative ones, can be provided by
our model, based on the DNS concept, and models [91-93], conceptually
close to it. In Fig.~\ref{fig:fig10}, as have already been mentioned in section~\ref{sec:dnsc}, the
potential energy of the DNS is shown as a function of the mass of the
light fragments produced in a QF process. Heavily indented due to the
shell effects, the potential energy has three global minima: at $M_L = 48,
82$ and $126$, the complementary heavy fragments being of $M_H =
208, 174$ and $130$ respectively. The first numbers of those two series
are easy to interpret: they are associated with doubly magic 
the $^{48}$Ca projectile and the $^{298}$Pb target. The second and shallower minimum
is associated with $^{82}$Ge ($Z = 32, N = 50$); in this case the fragment charge is
close to the magic number $Z = 28$ and the neutron number is exactly
equal to a magic neutron number, the complementary heavy fragment
being of a non-magic number. The third and deeper minimum is due to
the fact that both fragments are close to magic Sn.

\begin{figure}[p]
\includegraphics[width =5cm]{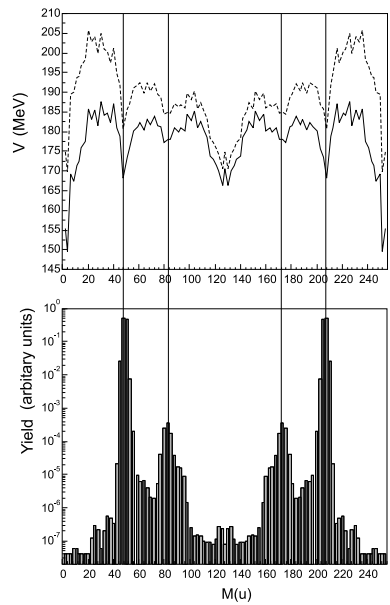}
\caption{\label{fig:fig11} Driving potential $V$ of the dinuclear system as a function of
fragment mass M. Shown below is a calculated mass spectrum for
the quasi-fission process occurring in this reaction.}
\end{figure}

\begin{figure}[p]
\includegraphics[width=7 cm]{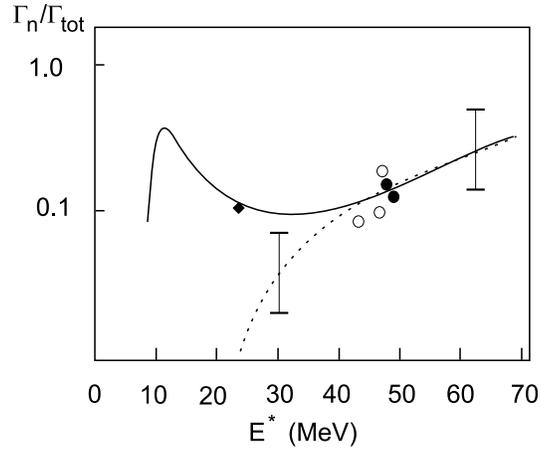}
\caption{\label{fig:fig12} Probability of neutron emission
$\Gamma _{\rm n}/\Gamma _{\rm tot}$ as a function of the $E^*$ of
the $^{256}$No compound nucleus. The result of calculations taking
account of the energy dependence of the shell correction is
indicated with a solid curve; the result of calculations on the basis
of the liquid-drop fission barrier is indicated with a dotted curve.
The figure is taken from ref. [46]; the experimental works from
which the experimental data were borrowed are referred to
therein.}
\end{figure}

\begin{figure}[p]
\includegraphics[width =7cm]{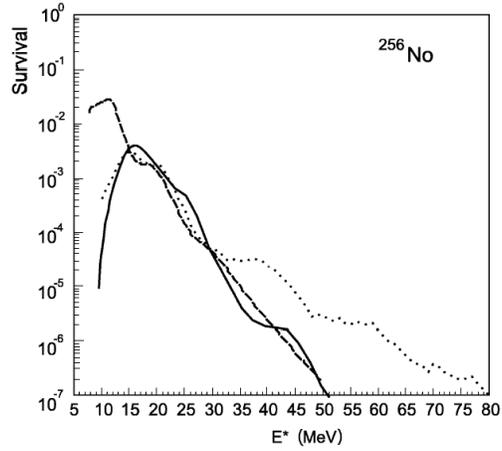}
\caption{\label{fig:fig13} Survivability of the $^{256}$No compound nucleus, calculated by three
different approaches, for statistical decay as a function of
excitation energy $E^*$. For details see the text.}
\end{figure}

\begin{figure}[p]
\includegraphics[angle=-90,width=12cm]{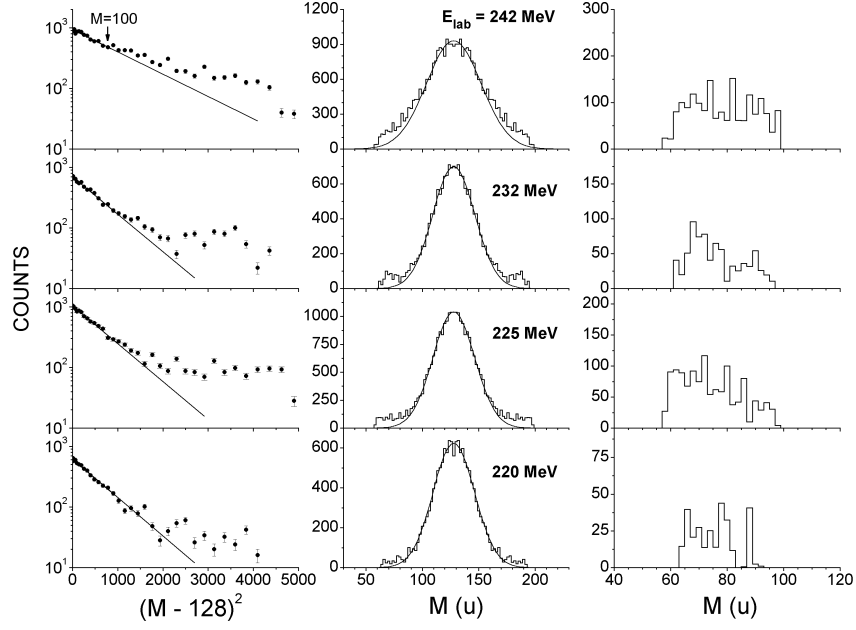}
\caption{\label{fig:fig14} At the left: Mass yields of fragments as a function of the
parameter $(M-128)^2$ (logarithmic scale) for the energies of $^{48}$Ca-ions 
$E_{\rm lab} = 220-242$ MeV. On such a scale, a Gaussian is a straight
line. At the center: Mass distributions of fragments (linear scale)
and fitting Gaussians, as at the left of the figure. At the right: The
separated quasi-fission shoulders as a function of the light fragments
mass.}
\end{figure}

\begin{figure}[p]
\includegraphics[width=8cm]{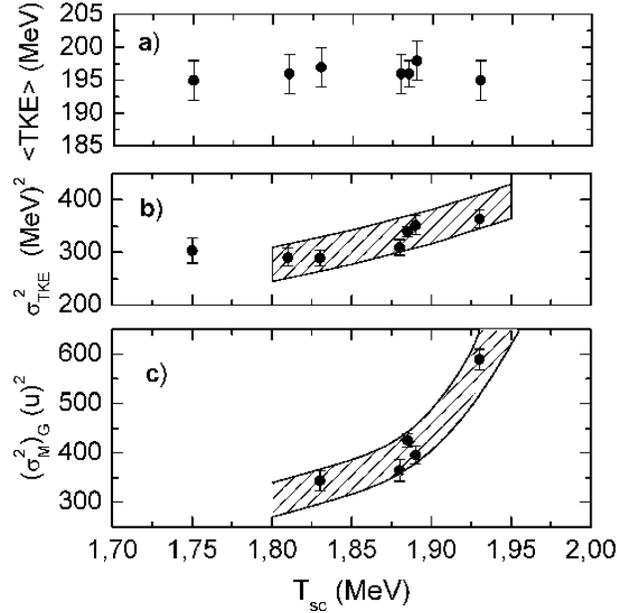}
\caption{\label{fig:fig15} The dependences of $\langle TKE \rangle$ (a),
$\sigma^2_{\rm TKE}$ and ($\sigma^2_M$)$_G$ on the nucleus
temperature at the scission point $T_{\rm sc}$ (21). The regions of $\sigma^2_{\rm TKE}$ and
($\sigma^2_M$)$_G$ characteristic of classical fission are shown by hatching
(see the text).}
\end{figure}

\begin{figure}[p]
\includegraphics[width=8cm]{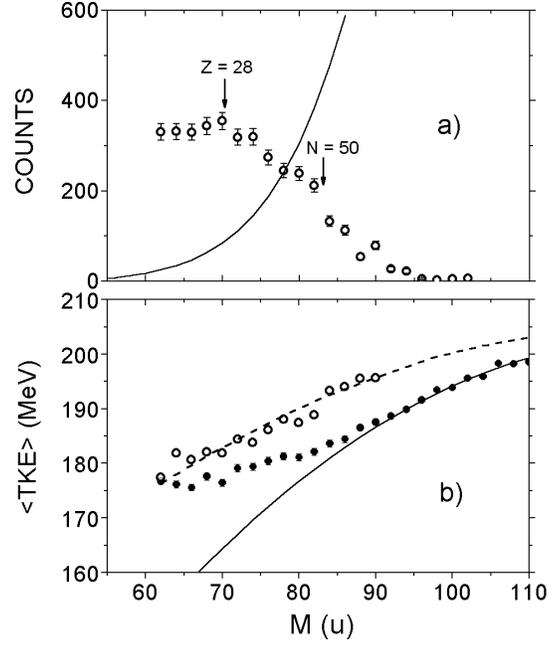}
\caption{\label{fig:fig16}  (a) Yield of the quasi-fission component for the light fragments
(open circles). Indicated with a solid curve is a portion of the
Gaussian from Fig.~\ref{fig:fig7}. The fragment masses with magical $Z$ and $N$
are indicated with arrows; the calculations were done by UCD.
 (b) M-dependences of the experimental $\langle TKE \rangle$ (black circles)
and the $TKE$ of the quasi-fission component (open circles)
obtained according to (26). The solid and dotted curves are the
fitting parabolic curves for those dependences according to Eq.~(27).}
\end{figure}

\begin{figure}[p]
\includegraphics[angle=270,width=10cm]{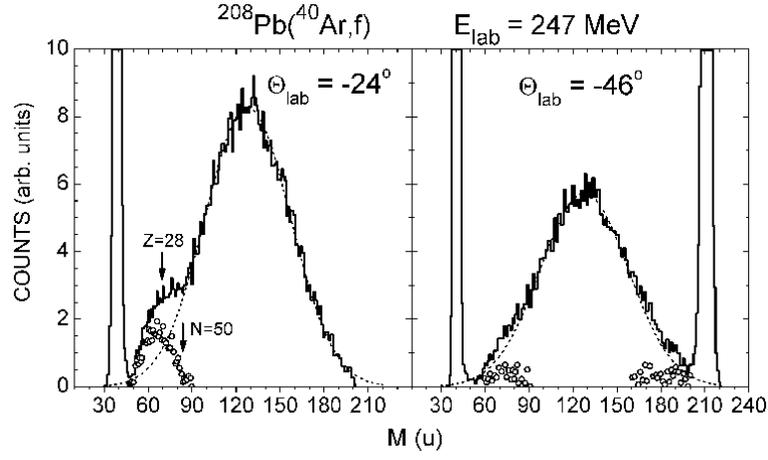}
\caption{\label{fig:fig17} Mass yields of fragments, presented as
a histogram, for the reaction $^{40}$Ar$+^{208}$Pb for two
registration angles from work [83]. The fitting Gaussians for FF
are indicated with dashed curves. The QF component is indicated
with open circles. The fragment masses with magical $Z$ and $N$,
calculated by UCD, are indicated with arrows.}
\end{figure}

At the bottom of Fig.~\ref{fig:fig11}, the mass distribution of QF products is shown;
it was calculated in the way described in section~\ref{sec:comp}. It 
 reflects completely the structure of the DNS potential energy. The
first peak, at $M_L \approx 48$, can in no way be distinguished in experiment
from the peak associated with deep inelastic transfer reactions, which is
in fact seen in Figs.~\ref{fig:fig6} , \ref{fig:fig7} and \ref{fig:fig8}.
The second peak with average mass $M_L \approx
82$ is associated with the quasi-fission shoulder, which was found
experimentally. The third peak in Fig.~\ref{fig:fig11}, at mass $\sim A/2$, is
characterized by a yield that is three orders of magnitude smaller than that
at mass $M_L \approx 82$. Although that peak is associated with the deepest
minimum of the potential energy it will be practically impossible to
observe this peak in experiment because the yield of FF is predominant.
This minimum of the DNS potential energy is populated slightly, which
is accounted for by the fact that the probability of the DNS breaking up
before reaching a symmetric configuration is high. The total 
capture-fission cross-section $\sigma _c$ in Fig.~\ref{fig:fig5} 
(indicated with a solid line) was
calculated taking account of both the FF cross-section $\sigma _{\rm FF}$ and
the whole sum of the mass yields associated with QF (this sum is shown
in Fig.~\ref{fig:fig11}), the peak at $M_L \approx 48$, which is responsible for a major part
of the calculated QF cross-section, being taken account of as well. To
consider the experimental MD, the contribution of QF to the total MD is
$15-20\%$ for the operating angular range of our spectroscope on
condition that only fission-like events are taken into account, as shown in
Fig.~\ref{fig:fig7}.

In our model as well as in models from [91, 92], the DNS excitation energy 
is counted from the DNS potential energy for each mass instead of the CN 
ground state. That is why the excitation energy of fission fragments, 
for example with $M = 82$, will be by about $20$~MeV lower than $E^*$ 
of the compound nucleus, since it is exactly this value by which the 
potential energy for this mass is ''lifted'' over the CN ground state. 
Thus, the fission fragment having this mass formed in the 
reaction $^{48}$Ca$+^{208}$Pb will be more cold than that formed in the FF process.

\begin{figure}[p]
\includegraphics[width=8cm]{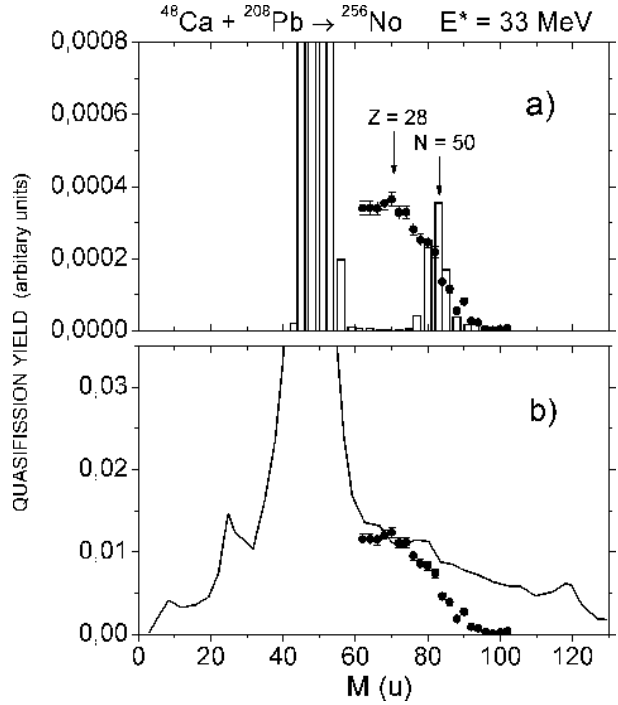}
\caption{\label{fig:fig18} (a) Comparison of the QF yields obtained
in experiment (circles) and calculated by our model (histogram).
With arrows, the same is indicated as 
in Fig~\ref{fig:fig16}(a). 
(b)Comparison of the QF yields obtained in experiment (circles) and
calculated in ref. [92] (solid curve).}
\end{figure}

In Fig.~\ref{fig:fig18}(a), the relative mass yields of light fragments for QF obtained
experimentally and calculated by our model are compared. It is clearly
visible that the experimental distribution is far wider than that predicted
theoretically and that in this case it lacks a clearly defined peak-like
shape, while, as seen from data on the fission of $^{248}$Fm (Fig.~\ref{fig:fig17}), the
experimental quasi-fission component is really peak-like in shape. That
the theoretical and experimental distributions for QF differ in width is
most likely to be connected to the fact that a statistical approach has
limited application in calculating such a QF characteristic as MD, since
according to works [5, 7, 8] the formation of fragment mass in QF is a
dynamical process with the mass relaxation time $\tau = 5.3 \times 10^{-21}$ c
[8]. Therefore our statistical calculations can be considered as a first
approximation that only predicts QF characteristics in a qualitative way.

In recent work [92] a dynamical transport model of QF has been
developed on the basis of the DNS concept. In this model, which takes
account of shell effects, a change in the mass (charge) asymmetry of DNS
is a result of diffusion. Fig.~\ref{fig:fig18}(b) shows the experimental and calculated
[92] relative yields  associated with QF for the reaction concerned. As
can be seen from that figure, the MD for QF calculated in the framework
of this model is a distribution that smoothly rolls off to become
symmetric and resembles the experimental distribution, there being minor
peculiarities due to the shell effects discussed above. This MD is far
wider than that that we found. In model [92], the QF process has a
characteristic time of $30-40 (\times 10^{-21})$ s, which is $3-8$ times larger
than that found in experiment [7, 8]. Such a time is typical of the classical
fission of heavy nuclei [22, 66, 67, 77, 83], and the angular distribution
for each mass will be symmetric about $90^\circ$, which contradicts
experimental data [7, 8, 48-51].

As a result, we can conclude that on the whole both our model of DNS
and the model from work [92] provide qualitatively correct predictions
about some properties of QF however these models need improving for
them to be capable of making quantitative predictions about the
characteristics of QF.
The theoretical approach realized in work [93] seems, in our opinion, promising.
At present, this approach, which
takes account of shell effects, is the most consistent in treating the fusion
and quasi-fission of heavy nuclei in reactions with massive ions.

\subsection{Properties of low-energy fission}
\subsubsection{Analysis of experimental results}

 It is now common knowledge that for spontaneous
fission, the MEDs of fragments for the superheavy nuclei beginning with $^{258}$Fm are
bimodal in structure [19, 94-97]. Typical for those nuclei are "anomalously"
narrow symmetric MDs with the high total kinetic energy of fragments of
up to $ \sim 235$~MeV.
At the same time, the MD of those nuclei has a component of much larger width but
with the lower of $TKE \sim 200$~MeV, which corresponds to the standard fission of those nuclei. The ratio
of the high-energy to the low-energy component (mode) significantly varies
with the nucleon composition, but the universal tendency is that for $Z$ fixed, there is
a sharp increase in the yield of the symmetric mode with high $TKE$ (see the
systematics of mass yields and $TKE$ distributions in works [96, 97]) as the
neutron number of both fragments approaches the $N = 82$ neutron shell. To compare,
for example, the MDs for chains of Fm and No isotopes in spontaneous fission,
the standard two-humped distribution transforms drastically to the clearly defined
narrow symmetric distribution as the neutron number of a fissioning nucleus changes
from $157$ to $158$ in the former case and from $154$ to $156$ in the latter case in spite
of the fact that in the $TKE$ distributions, the high-energy mode is dominant in the
case of $^{258}$Fm and the low-energy mode in the case of $^{258}$No [94, 95].

The properties of the MDs of fragments of the nuclei under discussion can change
rather drastically, for example, from an asymmetric distribution, in the case of
$^{256}$Fm(s, f) [97-99], in which the nucleus is not excited, to a symmetric
distribution, in the reaction $^{255}$Fm($n_{\rm th},f$),
the excitation $E^* = 6.4 $~MeV [98,100],
or from a very narrow symmetric distribution to a distribution also symmetric but
several times larger in width, as in the case of $^{258}$Fm(s,f) [94] and
$^{257}$Fm(n$_{\rm th}$, f)
[101, 102]. At the intermediate excitation ($E^* = 10-20$~MeV), the shell effects begin to damp
out, but not to such a degree that the modal structure be unobservable. For
superheavy nuclei, this is most clearly demonstrated in work [98],
in which it is shown that for the  already discussed $^{256}$Fm nucleus, the high-energetic
mode appears in the thermal-neutron-induced fission of $^{256}$Fm whereas it is absent in
the spontaneous fission of $^{256}$Fm. Its further behaviour is traced up to
$E^* = 17.5$~MeV, at which it also manifests itself, but not so clearly.

Recently, it was shown in our work [19] that bimodality is also exhibited by the
nucleus of $^{270}$Sg ($Z = 106$) excited up to $E^* = 28$~MeV. However, it should be noted
that this is a unique case since the breakup into exactly equal parts results in
fragments with the magic neutron number $N = 82$.

Theoretical calculations [103-111] made for superheavy nuclei
 showed that on the potential energy surface in the multidimensional
deformation space of fissioning nuclei, there exist at least two  [104, 105, 108-111] or
even three [103, 106, 107] fission valleys for that process to follow. One of the
valleys, named Super-Short by Brosa et al. (SS) [103, 104], enables those nuclei to
have near-magic numbers of neutrons and protons of $N \sim 82$, $Z \sim 50$ in both fragments.
For superheavy nuclei, the SS valley was found in calculations [105, 108] to persist
up to $^{270,272}$Hs ($Z = 108$) and even up
to $^{278}110$ [104] and $^{290}110$ [106].

\begin{figure}[p]
\includegraphics[width=5cm]{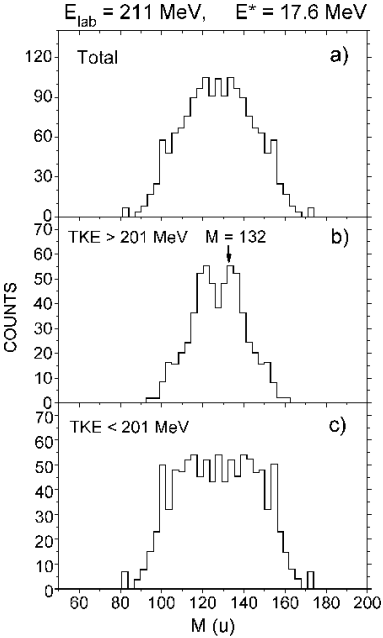}
\caption{\label{fig:fig19} Fragment mass distributions for
$E_{\rm lab}=211$~MeV ($E^*=17.6$~MeV):
total distribution (a),  and differential distributions for
$TKE > 201$~MeV(b) and $TKE < 201$~MeV(c).}
\end{figure}

Now let us consider the properties of the MEDs for low-energy fission of $^{256}$No.
Fig.~\ref{fig:fig19} presents the total (Fig.~\ref{fig:fig19}(a)) and differential distributions of fragment masses for
$TKE > 201$~MeV(Fig.~\ref{fig:fig19}(b)) and $TKE<201$~MeV(Fig.~\ref{fig:fig19}(c)) 
at the lowest energy $E_{\rm lab}=211$~MeV ($E^* = 17.6$~MeV). For
the high $TKE$, it is clearly visible that the MDs are two-humped in structure in
the region of the mass of heavy fragments $M_H \cong 131-135$, which is characteristic
of the high $TKE$ in the fission of all actinides. For superheavy nuclei, the shape of the
distribution very much resembles the MDs of fragments in the spontaneous fission
of $^{259}$Lr $(Z = 103)$ [112]. For the low $TKE < 201$~MeV, seen in Fig.~\ref{fig:fig19}(c) is a
wide and flat distribution held by the masses $M_H \cong 140-150$. Such shape of the MD
low-energy component is also typical of the fission of superheavy nuclei;
for example, the spontaneous fission of $^{259}$Md [94] or $^{260}$Md [113], in which the
low-energy component even has a slight valley for the symmetric masses. Thus, we
see that at $E^* = 17.6$~MeV, the MD of
fragments for $^{256}$No consists of two components at least.

\begin{figure}[p]
\includegraphics[width=8cm]{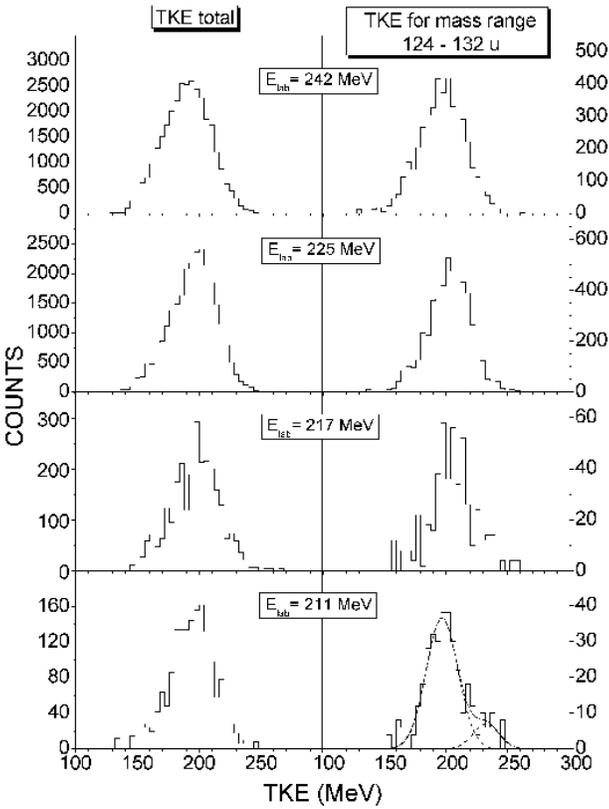}
\caption{\label{fig:fig20} $TKE$ distributions for four values
of the energy of $^{48}$Ca-projectiles.
At the left: Total distributions; at the right: Differential
distributions for the symmetrical mass region $M = 124-132$. For the
figure on the lower right, fitting Gaussians are shown for two
modes.}
\end{figure}

Let us consider the energy distributions (EDs) of fragments for $^{256}$No. Presented at
the left of Fig.~\ref{fig:fig20} are the distributions $TKE$ for four values of the energy
of  $^{48}$Ca-projectiles. For the two lowest energies,
attention is attracted by the fact that the
bottom of the distributions is larger in width. Shown at the right of
Fig.~\ref{fig:fig20} are the
EDs of fragments for the same energies of $^{48}$Ca ions, but only for the symmetric masses
in the range $ M = 124-132$, that is, for those masses where a two-humped MD is
experimentally observed (Fig.~\ref{fig:fig19}). For the largest
energies $E_{\rm  lab} = 242$ and $225$~MeV,
those distributions are practically not different from the total distributions in any
way. But for the two smallest energies, especially
for $E_{\rm  lab} = 211$~MeV ($E^* = 17.6$~MeV),
two $TKE$ components are distinctly visible. One of them is a dominant
low-energy component with $\langle TKE \rangle _{\rm low }\sim 200$~MeV,
and the other is a high-energy one with a
$\langle TKE\rangle _{\rm high} \sim 233$~MeV. Those $TKEs$ as well as the mass yields
are characteristic respectively of the standard and the SS mode in the spontaneous
fission of superheavy nuclei [94-97]. So, the experiment shows the fission of
excited $^{256}$No to be bimodal, as demonstrated both by the MD and ED, though the
contribution of the SS mode on the whole is slight.

We will point out especially that we think the SS mode to be inherent in
the classical low energy fission of $^{256}$No, though, as discussed
above, from the theoretical standpoint such fragments may be produced
in quasi-fusion as well, their yield being smaller as compared with
fragments of $M = 60-90$ (Figs.~\ref{fig:fig11} and \ref{fig:fig18}). However, 
to return to Fig.~\ref{fig:fig8}, in
which the two-dimensional matrix $(M-TKE)$ for $E_{\rm lab}=211$~MeV
is given, the matrix only has isolated events for the region of $M$ and $TKE$ essential to
 the quasi-fission shoulders. So there should be no QF events
associated with masses $M = 120-135$ at all and our assumption that the
SS mode occurs in a classical FF process is, in our opinion, irrefutable.

\begin{figure}[p]
\includegraphics[width=5cm]{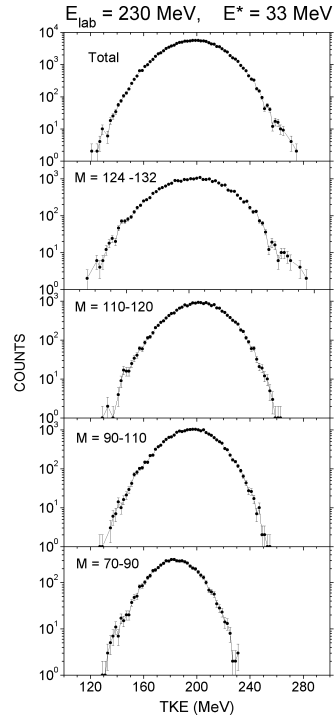}
\caption{\label{fig:fig21} Total and differential $TKE$ distributions for various mass ranges
of fragments at the energy $E_{\rm lab}=230$~MeV.}
\end{figure}

Now let's return to Fig.~\ref{fig:fig7} from which it can be seen that the dependence
$\sigma^2_{\rm TKE} (M)$ has a clearly noticeable small yet statistically
important peak in the symmetric region. An irregularity in the
dependence $\sigma^2_{\rm TKE} (M)$ always indicates that in the
mass region in which this irregularity is observed, the MED of fragments
is of complex structure. To account for this fact, the fragment yields as a function of $TKE$  
for the total mass region and different mass ranges 
are shown in Fig.~\ref{fig:fig21}. It can be seen that in the distributions for the total 
mass region and for the symmetric mass range $M = 124-132$, there is a 
high energy component similar to that at an energy of  $211$~MeV but 
with a smaller yield. This component becomes noticeable at very high
$TKE > 245$~MeV because it is covered by the main dome curve of
symmetric fission. Nothing of the kind is observed for the other mass
ranges.

\begin{figure}[p]
\includegraphics[width=5cm]{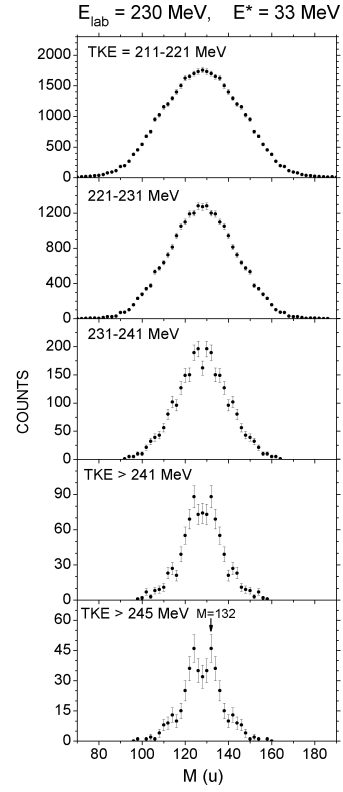}
\caption{\label{fig:fig22} Fragment mass distribution for different $TKE$ ranges for the beam
energy $E_{\rm lab}=230$~MeV.}
\end{figure}

Fig.~\ref{fig:fig22} shows the MDs of fragments for various high energy ranges of
$TKE$. For the ranges of $TKE=211-221$~MeV and $221-231$~MeV, the
MDs are near-Gaussian in shape though their peaks are seen to be
narrow. For the highest $TKE  > 241$ and $245$, a two-humped MD is
seen with a peak due to the heavy fragment of $M = 132$, there being
obviously two components: a two-humped narrow component and a wide
pediment-like component. So at $E^* = 33$~MeV, as at $E^* =
17.6$~MeV, the SS fission mode is seen, but its contribution to the total
distribution is small. At those values of $E^*$, the low-energy mode is already
close to the liquid-drop mode in properties.

Let us make a qualitative analysis of the picture obtained. To follow the traditional
assumption that the charge of a fragment is proportional to its mass, for the
region $M_H \cong 131-135$ its charge will be $Z_H = 52-54$, the neutron number being
$N_H = 79-81$. That is, the $Z$ of the heavy fragment is close to a magic number ($Z = 50$),
but slightly larger than it, and the neutron number is also a near-magic number
($N = 82$), but slightly smaller than it. For the complementary light fragment,
the yield peak lies in the mass region $M_L = 121-125$; if so, its charge will
be $Z_L = 48-50$, and $N_L = 73-75$. Then the light fragment has charge close to magic
charge and a neutron number far smaller than a magic number. So the SS mode is due
to the proximity of the nucleon composition both by $N$ and by $Z$ to a magic number for
the heavy fragment and only by $Z$ for the light. Since the total number of neutrons is
nevertheless not enough to close the spherical shells in both fragments, this mode is
not dominant for $^{256}$No and manifests itself only slightly.

According to various theoretical estimates [114-116], the fission barrier for cold
$^{256}$No $E_f = 6.4-8$~MeV.
 In recent experimental [117] and theoretical [118] works, it has been
demonstrated that $^{254}$No produced in a $^{208}$Pb ($^{48}$Ca, $2n$)
reaction has a fission barrier $E_{f}  > 5$~MeV for a nuclear spin of
$12 \hbar$ till $22 \hbar$. Therefore in our case the excitation energy at the
saddle point will be
$E_{\rm sp} = 10-12$~MeV for $E_{\rm lab}=211$~MeV and $E_{\rm sp} = 22-24$~MeV for
$E_{\rm lab}=230$~MeV with taking into account $0.3$ pre-scission neutrons.
This is rather low excitation, and shell effects, responsible for
the production of fission valleys, certain to play a key role in the formation of
the MEDs of fragments, which reflects on their structural peculiarities in
Fig.~\ref{fig:fig19}-\ref{fig:fig22}.

\subsubsection{Theoretical calculations}

The theoretical calculations of the $^{256}$No potential energy landscape  in relation to the
distance between the centeres of the future fragments and their elongations
are presented in work [105]. On that landscape, an SS-valley is clearly visible.
The valleys ramify into a standard valley and the SS-valley from the second minimum,
but the external potential barrier in the SS-valley is higher than that in the
standard valley, and it is more advantageous for a nucleus, while fissioning, to
overcome the traditional standard barrier, whose penetrability is higher than that
of the barrier in the SS-valley. Therefore the SS-mode is practically unobservable
in spontaneous fission [95, 97, 119]. In induced fission, as excitation energy
increases, firstly, the values of barriers decrease in magnitude [120] (shell
effects damp out); secondly, the potential energy landscape gets smoother [120],
which decreases the difference in the heights of the barriers in the different
valleys; thirdly, excitation of the nucleus allows an SS-valley of higher energy
to be populated, though with lower probability than a standard valley.

This phenomenon, i.e the manifestation of some mode
which was not observed in the spontaneous or near-barrier fission
and its appearance at a certain  excitation of the nucleus, has
been already observed  for different nuclei. In works [55, 121] for $^{213}$At  the
appearance of the standard symmetric mode was observed  against the background of
the dominant symmetric mode. In [122, 123] for the region of heavy isotopes Ra-Ac,
 on the contrary, the appearance  of the symmetric mode was observed at a dominance
of the standard mode; in [98, 100] for $^{256}$Fm, already mentioned, the appearance of
the SS mode was observed at a  dominance of the standard mode. All those data, as
with $^{256}$No, can only be accounted for in terms of independent fission modes
arising from the fact that there are valleys on the surface of the potential
energy in the multidimensional deformation space of the fissioning nucleus.

In this paper we present new multi-parametric calculations of the 
$^{256}$No potential energy near the scission point as a function of mass asymmetry. 
The calculations were made with the use of the Strutinsky shell correction method. 
The surface of the nucleus was described in parametrization based on 
Cassinian ovals [107], where parameter  $\alpha$ characterizes the nucleus 
elongation and parameter $\alpha_3$ and others odd number parameters describe  
the asymmetry of the figure. Deformations of a higher order of up to   
$\alpha_{31}$ were taken into account in the process of minimization.   

\begin{figure}[p]
\includegraphics[width=10cm]{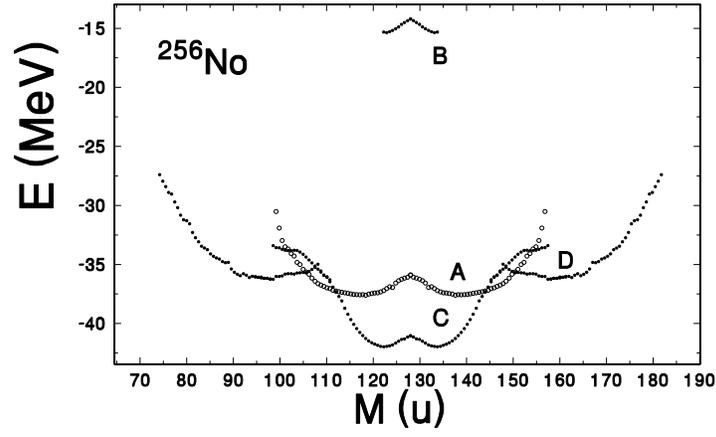}
\caption{\label{fig:fig23} The  $^{256}$No  fission  valleys  near   the
scission point as a function of the fragment mass.}
\end{figure}

Figure~\ref{fig:fig23} shows the potential (deformation) energy of $^{256}$No near  
the scission point at the fixed neck radius $R_{\rm neck} = 0.3 R_0$  
($R_0$ is the radius of the spherical nucleus of the same volume).  
The neck radius is close to the critical one as the fissioning  
nucleus goes over from the fission valley to that of the separated 
fragment [124]. The calculations have shown that the process of fission 
near the scission point evolves in four more or less independent valleys, 
indicated by letters A, B, C and D in Fig.~\ref{fig:fig23}. It looks like the valleys 
A, C and D intersect in the considered coordinates. This happens because 
of the multidimensionality of our calculations, i.e. the valleys lying 
in the area of intersection are characterized by different deformations 
of a higher order and are separated by the potential barriers. It means 
that the fragments having the same mass (for example $M_H = 148$) can be 
produced in three different valleys and they will have $3$ different 
deformation sets, and, as consequence $3$ sets of $\langle TKE \rangle$, as it was 
clearly demonstrated in the experiments [55] using $^{210}$Po and $^{213}$At. 
In fact this is the essence of the multi-modal approach to the 
fission process. The question is with what statistical weight every 
specific mode will contribute into the total $(M-TKE)$ distribution? 
As mentioned above, this contribution will be determined by the ratio of 
heights and penetrations of the fission barriers in every valley.

\begin{figure}[p]
\includegraphics[width=12cm]{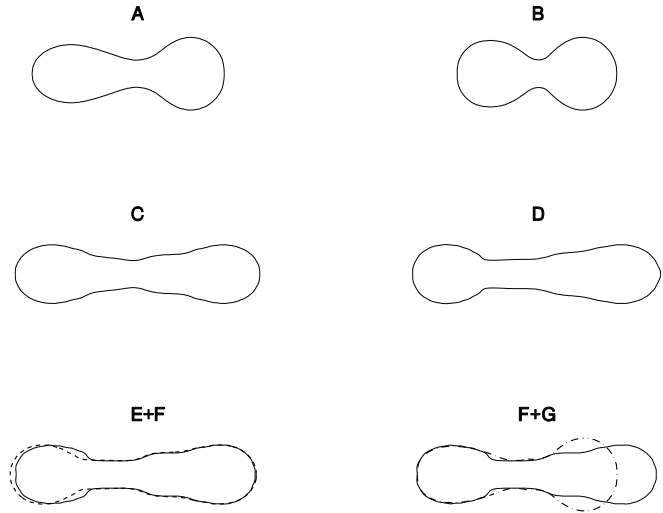}
\caption{\label{fig:fig24} 
Shapes of the fragments at the bottom of the valleys shown 
in Fig.~\ref{fig:fig23}, and comparison of the shapes at $M = 148$. Shapes in points E (dashed), 
F (solid) and G (dash-dotted) stem from the bottoms of the valley C, D and A, 
respectively, up to the intersection points.}
\end{figure}

We have not yet finished the calculations for the $^{256}$No fission valleys, 
the results will shortly be published elsewhere. However according to the 
preliminary data, the valley A is the first candidate for the population 
in the fission process. As one can see in Fig.~\ref{fig:fig23}, the bottom of this 
valley corresponds to the fragments with masses $M \approx 135-145$, and a more 
probable nuclear shape near the scission point is shown in Fig.~\ref{fig:fig24}(A). 
In this case the shape of the heavy fragment is close to the spherical one, 
and the deformation of the en-tire fissioning nucleus is quite compact. 
It corresponds to the fission of $^{256}$No in a standard asymmetric valley 
and this is in excellent agreement with the experimental data on the 
spontaneous fission of $^{256}$No (fig.~8(a) from [119]), as well as with our results.
Comparing the potential energy in different valleys, valley B is 
situated quite high (see Fig.~\ref{fig:fig23}), and its more probable nuclear 
shape is presented in Fig.~\ref{fig:fig24}(B). It is the Super-Short (SS) 
valley with very compact shapes of both fragments and mass of 
heavy fragment $M_H \approx 134$, that also agrees well with our experimental 
results. However, as the calculations show, such a compact shape is 
extremely unstable with respect to small changes in the deformation, 
and the nucleus tends to take shape A, since the light fragment 
is far from being magic in the neutron number. Therefore, the evolution 
of the fission process along valley B is unlikely, and the yield of the 
SS-mode fragments should be small, as we have observed in the experiment.

In valleys C and D (Fig.~\ref{fig:fig23}) the fragments are strongly deformed, and the 
most probable shapes, corresponding to the division into masses $122/134$ 
and 95/161, are shown in Fig.~\ref{fig:fig24}(C) and (D). Although valley C has the minimum 
of the potential energy corresponding to mass $M_H \approx 134$, it is not connected 
with the spherical shells $Z=50$ and $N=82$. The fragment shapes are so 
strongly deformed, that by their shapes they are close to those 
obtained in the LDM [125], and all the properties 
of the valley are mainly due to the liquid-drop component of the potential 
energy, but ''distorted'' by the strongly deformed shells, as it 
was shown for the pre-actinide nucleus in [126]. In Brosa's terminology, 
this valley as well as the corresponding mode is called the 
Super-Long (SL) mode. In spontaneous and low-energy fission this mode 
practically does not manifest itself. With increasing excitation energy, 
its properties become similar to those in the LDM, as it happens in our 
experiment (Figs.~\ref{fig:fig6} and \ref{fig:fig7})

Valley D is formed by the light fragment cluster, whose properties are 
close to those of the third standard mode S3 recently found in ref.~[80], 
in which the neutron number in the light fragment $N \approx 52$ [80] or 
slightly more [127]. This mode is a very asymmetric one and difficult 
to observe in the experimental mass distributions. In our case it 
is just the statistical accuracy that is lacking.

Finally, the shapes of the fissioning nucleus for mass $M_H = 148$, 
produced in three different valleys C, D and A are compared 
in Fig.~\ref{fig:fig24}(E+F) and Fig.~\ref{fig:fig24}(F+G). 
This presents a good example of the fission modality.
Thus, our theoretical studies explain clearly the main properties 
and the regularities of the low-energy fission of $^{256}$No 
investigated experimentally.

\section{ Conclusions}

For the reaction $^{48}$Ca $+ ^{208}$Pb$ \to ^{256}$No, the capture-fission
cross-sections and the mass energy distributions have been measured for
the energies of $^{48}$Ca-projectiles respectively from $206$ up to
$242$~MeV and from $211$ up to $242$~MeV with the help of the
CORSET double-arm time-of-flight spectrometer. The MED of fragments
for heated fission has been shown to consist of two components. One of
the components is classical fusion component due to the symmetric
fission of the $^{256}$No compound nucleus, and the other component,
which manifests itself as ''quasi-fission shoulders'', is associated with a 
quasi-fission process. The quasi-fission shoulders, which embrace light fragments
of a mass of $\sim 60-90$, show increased $TKE$ as compared with that
expected for a FF process. The conclusion has been drawn, in agreement
with work [54], that in QF, the spherical shells with $Z = 28$ and $N =
50$ play a key role. This conclusion is supported by results from work
[83]. The MD of fragments for the $^{40}$Ar $+ ^{208}$Pb$ \to ^{248}$Fm
reaction obtained in this work has marked quasi-fission shoulders whose
characteristics are close to those of the quasi-fission shoulders associated with
the reaction $^{48}$Ca $+ ^{208}$Pb $\to ^{256}$No. The properties of
MED of fragments have also been demonstrated to formally fit the
standard hypothesis about two independent fission modes; in this case,
the two modes are only a normal FF process and a QF process, which
exhibit their features.

A version of the Dinuclear System model has been developed,
which has enabled the experimental energy dependences of the 
capture-fission cross-sections and the $xn$-channels of the reaction concerned to
be described. This version of the DNS model, which takes account of
shell effects, provides qualitatively correct predictions about the mass
distribution for QF.

A comparison of the mass distributions for QF measured in experiment
and calculated by the dynamical transport model [92] shows that this
model also provides qualitatively correct predictions. However both
models need improving to a large measure to provide quantitative
predictions.

It has been found that at low excitation energies, there exists a high
energy Super-Short mode of classical fission in the mass region of heavy
fragments $M = 130-135$, the $TKE \approx 233$~MeV, but the yield
of this mode is small. Its properties can be accounted for with the help of
the classical modal approach, which is based on the fact that the potential
energy surface in the deformation space of the fissioning nucleus has
valley structure.

\acknowledgments
We are pleased to acknowledge help from Dr.~V.~S.~Salamatin in data
processing and advices from Prof.~V.~V.~Volkov in conducting theoretical
investigations. One of us, (C.~E.~A), thanks Prof. V.~I.~Zagrebaev for his
program for calculating driving potential and numerous helpful and
fruitful discussions. P.~E.~V. thanks Prof.~F.~Hanappe for helpful discussions.
This work was supported by  the Russian Foundation for Basic
Research under Grant No. 03-02-16779, and by INTAS grants YS 2002-261 and partly No. 00-655.


\begin{table}[h]
\caption{\label{tab:table1} Characteristics of the reactions and
some experimental results.}
\begin{ruledtabular} 
\begin{tabular}{cccccccc}
$E_{\rm lab}$& $E^*$ &$ \sigma_{c}$ &$\langle TKE \rangle$&
$\sigma^2_{TKE}$ &$(\sigma^2_M)_G $& $\langle\nu _{\rm pre}\rangle$&$ T_{\rm cs}$\\
(MeV)& (MeV) & (mb) & (MeV) & $(MeV)^2$& $(u)^2$  & & (MeV)\\
\hline
$242$ & $42.8$ & $480 \pm 35$& $195 \pm 3$& $364 \pm 17$& $590 \pm 21$& $1.0$& $1.93$\\
$232$ & $34.7$ & $390 \pm 30$& $198 \pm 3$& $352 \pm 18$& $396 \pm 18$& $0.4$& $1.89$\\
$230$ & $33.1$ &             & $196 \pm 2$& $340 \pm 10$& $426 \pm 15$& $0.3$& $1.885$\\
$225$ & $29.0$ & $240 \pm 22$& $196 \pm 3$& $310 \pm 15$& $365 \pm 22$& $0  $& $1.88$\\
$220$ & $25.0$ & $84 \pm  8 $& $197 \pm 3$& $290 \pm 15$& $344 \pm 20$& $0  $& $1.83$\\
$217$ & $22.5$ &             & $196 \pm 3$& $291 \pm 17$&             & $0  $& $1.81$\\
$211$ & $17.6$ & $8\pm 2    $& $195 \pm 3$& $304 \pm 24$&             & $0  $& $1.75$\\
$206$ & $13.6$ & $0.045 \pm 0.031$ &&&& \\
\end{tabular}
\end{ruledtabular}
\end{table}


\end{document}